\documentclass[12pt, draftclsnofoot, onecolumn]{IEEEtran}
\usepackage{amsmath, graphics,amssymb,epsfig, subfigure}
\usepackage{graphicx,color,rotating}
\usepackage{multirow}
\usepackage{cite}
\usepackage{color,xcolor}
\usepackage{algorithm}
\usepackage{algorithmic}
\usepackage{cases}
\usepackage{booktabs}
\usepackage{soul}

\newtheorem{theorem}{Theorem}

\newtheorem{remark}[theorem]{Remark}

\newcommand{\be}{\begin{equation}}

\newcommand{\ee}{\end{equation}}
\newcommand{\bea}{\begin{eqnarray}}
\newcommand{\eea}{\end{eqnarray}}
\newcommand{\bdp}{\begin{displaymath}}
\newcommand{\edp}{\end{displaymath}}
\makeatletter
\def\hlinewd#1{%
\noalign{\ifnum0=`}\fi\hrule \@height #1 \futurelet \reserved@a\@xhline}

\DeclareMathOperator*{\argmax}{arg\,max}

\makeatother

\begin{document}
\title{Deep Learning for Multi-User MIMO Systems: Joint Design of Pilot, Limited Feedback, and Precoding}
\author{\IEEEauthorblockN{\normalsize{Jeonghyeon Jang, Hoon Lee, \textit{Member}, \textit{IEEE}, \\ Il-Min Kim, \textit{Senior Member}, \textit{IEEE}, and Inkyu Lee, \textit{Fellow}, \textit{IEEE}}} \\

\thanks{This work was supported in part by the National Research Foundation of Korea (NRF) Grant funded by the Korea Government (MSIT) (No. 2022R1A5A1027646, 2021R1I1A3054575), in part by Natural Sciences and Engineering Research Council of Canada (NSERC). This article was presented in part at the IEEE Region 10 Symposium 2021, Jeju, South Korea, August 2021 \cite{Jang:21}. \textit{(Corresponding authors: Hoon Lee; Inkyu Lee)}

J. Jang, and I. Lee are with the School of Electrical Engineering, Korea University, Seoul 02841, South Korea (e-mail: $\{$march$\_$19, inkyu$\}$@korea.ac.kr).

H. Lee is with the Department of Smart Robot Convergence and Application Engineering and the Department of Information and Communications Engineering, Pukyong National University, Busan 48513, South Korea (e-mail: hlee@pknu.ac.kr).

I.-M. Kim is with the Department of Electrical and Computer Engineering, Queen's University, Kingston, ON, K7L 3N6, Canada (e-mail: ilmin.kim@queensu.ca).}
}
\maketitle \thispagestyle{empty}


\begin{abstract}
In conventional multi-user multiple-input multiple-output (MU-MIMO) systems with frequency division duplexing (FDD), channel acquisition and precoder optimization processes have been designed separately although they are highly coupled. This paper studies an end-to-end design of downlink MU-MIMO systems which include pilot sequences, limited feedback, and precoding. To address this problem, we propose a novel deep learning (DL) framework which jointly optimizes the feedback information generation at users and the precoder design at a base station (BS). Each procedure in the MU-MIMO systems is replaced by intelligently designed multiple deep neural networks (DNN) units. At the BS, a neural network generates pilot sequences and helps the users obtain accurate channel state information. At each user, the channel feedback operation is carried out in a distributed manner by an individual user DNN. Then, another BS DNN collects feedback information from the users and determines the MIMO precoding matrices. A joint training algorithm is proposed to optimize all DNN units in an end-to-end manner. In addition, a training strategy which can avoid retraining for different network sizes for a scalable design is proposed. Numerical results demonstrate the effectiveness of the proposed DL framework compared to classical optimization techniques and other conventional DNN schemes.
\end{abstract}



\section{Introduction}
Multi-antenna techniques have been regarded as key enablers for improving spectral efficiency in wireless systems \cite{Spencer:04, Love:08, Peel:05, Choi:04, Christensen:08, Shi:11, Jindal:06, Hong:22}. To implement multiple-input multiple-output (MIMO) precoding methods, it is essential to achieve closed-loop communication where a base station (BS) can compute appropriate precoding matrices based on channel state information (CSI) obtained at users. Channel acquisition strategies of the BS depends on duplexing systems. In time division duplexing (TDD), the CSI can be straightforwardly estimated at the BS through channel reciprocity. In contrast, frequency division duplexing (FDD) systems rely on limited feedback procedures \cite{Love:08, Jindal:06, Ravindran:08}, which needs additional step of quantizing the CSI at each user.

In the case of the FDD, the BS first broadcasts pilot sequences through downlink channels and users execute channel estimation processes. Using the received pilot sequences, the users first individually estimate their channels and transmit the resulting discrete CSI information back to the BS. Since neither the channel estimation nor the feedback is perfect, errors accumulate as the channel acquisition procedure progresses, which deteriorates the system performance. This becomes even more severe in multi-user (MU) MIMO configurations where the CSI at the transmitter (CSIT) uncertainty due to imperfect channel estimation and limited feedback exacerbates inter-user interference. Therefore, the FDD MU-MIMO systems require a careful design of precoders of all users considering CSIT errors. Particularly, such designs entail fundamental challenges in developing a joint optimization of the channel acquisition process including the pilot, the limited feedback and the precoder design, because separately designed precoder and channel acquisition mechanisms do not effectively mitigate the inter-user interference.

In traditional MU-MIMO systems, however, it is highly difficult to perform such joint optimization of the channel acquisition and the precoder computations. Even with perfect CSIT, a precoder design is not a trivial task due to its nonconvexity. There have been intensive efforts on solving the nonconvex MU-MIMO optimization problems \cite{Peel:05, Choi:04, Christensen:08, Shi:11, Moon:14, Sung:09, Zhang:09, Dabbagh:08, Choi:20, Vucic:09, Bogale:10, Fritzsche:13, Dai:16, Jindal:06, Ravindran:08}. A popular approach is the weighted minimum mean square error (WMMSE) algorithm \cite{Christensen:08, Shi:11} which identifies efficient precoding and decoding matrices through iterative methods. It has been reported that the WMMSE algorithm achieves locally optimum performance for the sum-rate maximization problem in MU-MIMO systems with perfect CSIT. However, the computational complexity for WMMSE is generally high, especially in high signal-to-noise ratio (SNR). Furthermore, the precoders were designed in \cite{Peel:05, Choi:04, Christensen:08, Shi:11} assuming perfect CSIT, i.e., the CSIT uncertainty is not considered.

In the presence of imperfect CSIT, precoder optimization techniques have been studied in the literature \cite{Zhang:09, Dabbagh:08, Choi:20, Vucic:09, Bogale:10, Fritzsche:13, Dai:16} and the designed precoders were demonstrated to be robust to CSIT errors to some extent. Nevertheless, such prior works cannot fully overcome the performance saturation issue at high SNR regime (i.e., the sum-rate performance does not improve any further beyond certain SNR values), which is often observed in limited feedback systems \cite{Jindal:06, Ravindran:08}. This is due to the limitations of conventional precoder optimization methods, which cannot jointly optimize the pilot sequences and the feedback which are closely related to CSIT errors. To address this issue, a new precoder optimization technique which includes a joint design of pilot sequences and CSI feedback is necessary.

This paper proposes a novel deep learning (DL) method that jointly optimizes the pilot sequences, CSI feedback mechanism, and the precoder for the FDD MU-MIMO systems through data-driven training algorithms. In the literature, there have been recent researches on DL approaches that solve multi-antenna communication problems \cite{Kim:20, Huang:20, Xia:19, Hu:20, Jang:20, Sohrabi:21, Kong:21}. Deep neural networks (DNNs) were employed in \cite{Kim:20, Huang:20, Xia:19} to design beamforming vectors in multi-user multiple-input single-output (MU-MISO) networks with perfect CSIT. It has been shown that DNN based MU-MISO systems achieve almost identical performance as the classical beamforming algorithms with reduced computational complexity. Nevertheless, these fully-connected DNNs have not been effective in the MU-MIMO case which is much more challenging than the MU-MISO case. The MU-MIMO case was studied in \cite{Hu:20}, in which the unfolding techniques were proposed by employing a layered structure of DNN modules. This facilitates reliable training of deep architectures, thereby improving the sum-rate performance compared to a naive DNN with only dense layers. However, the deep forward pass method in \cite{Hu:20} which involves a large number of iteration steps of the WMMSE algorithm requires high computational complexity as the number of DNN modules increases. Furthermore, the achieved performance still exhibits a non-negligible loss compared to the WMMSE algorithm. Moreover, the works in \cite{Kim:20, Huang:20, Xia:19, Hu:20} focused only on the transmitter optimization assuming perfect CSIT, and thus any joint design of the pilot sequences and the limited feedback along with the precoder was not studied.

Meanwhile, there have been recent DL researches on an end-to-end joint optimization of the channel acquisition and a precoder design \cite{Jang:20, Sohrabi:21, Kong:21}. The FDD single-user MIMO system was considered in \cite{Jang:20} by employing DNNs. The performance is fairly improved compared to the classical communication strategies that separately design channel estimation and quantization processes. This method has been extended to the FDD MU-MISO network \cite{Sohrabi:21}, which jointly determines the pilot sequences, limited feedback, and beamforming vectors via cascaded DNN units. The authors in \cite{Kong:21} have considered a joint design of the limited feedback and the precoder for the FDD MU-MIMO systems. However, it is still unclear whether the DL techniques proposed in \cite{Jang:20, Sohrabi:21, Kong:21} can perform well for the MU-MIMO scenario even for the perfect CSIT case, not to mention the practical imperfect CSIT case. Furthermore, in \cite{Sohrabi:21}, the feasibility of a fully-connected DNN structure in MU-MIMO cases has not been shown.

To the best of our knowledge, there has been no work which jointly designs the channel acquisition at the BS and the precoder design in MU-MIMO systems for either prefect or imperfect CSIT case. Unfortunately, such a joint design is very challenging even for the perfect CSIT case, because it is highly nonlinear and nonconvex. In our work, to tackle this difficult problem, we propose a new DL framework for FDD MU-MIMO networks which jointly optimizes the pilot sequence design, the channel feedback, and the precoder design. Each of these components is modeled as an individual DNN unit. First, assuming perfect CSIT, we develop a novel DNN at the BS, called the BS DNN, which determines the MIMO precoding matrices. Specifically, we propose a multi-stage learning policy which reduces online computational complexity by shifting online operations of DNNs to an offline training domain. In the proposed scheme, a single DNN module is trained at the BS with a stage-wise training loss function which is computed by the BS DNN trained at the previous stage. As the stages progress, the performance of the BS DNN is enhanced without requiring additional DNN employments.

Next, for the case of imperfect CSIT, we extend the BS DNN to a joint design of pilots, limited feedback, and the precoder. To this end, we introduce neural network (NN) modules which can be readily connected to the BS DNN to characterize the channel acquisition process of MU-MIMO networks. Specifically, we propose to use a NN for pilot design which allows users to obtain accurate CSI. We also employ a separate DNN for each user to perform distributed CSI quantization based on the received pilot signals, which is called the neural vector quantization. Finally, the BS DNN is refined to utilize the quantized CSI for the precoder design by implementing dequantizer DNNs. Then, all NN modules are jointly trained to improve the sum-rate performance in an offline manner. Additionally, we propose a scalable training method for an arbitrary number of users, which requires no retraining for different number of users. Numerical results demonstrate the following two main achievements. First, for the case of perfect CSIT, thanks to the proposed multi-stage training, the computational complexity of the BS DNN is significantly reduced compared to conventional schemes. Even with much reduced complexity, the sum-rate performance of the proposed scheme is better than the existing DL approaches and is quite close to the performance of the WMMSE. Second, for the imperfect CSIT, unlike conventional schemes our proposed scheme shows monotonically increasing sum-rate performance as SNR grows. This demonstrates that the proposed scheme can overcome the performance saturation problem that has been faced by other conventional schemes.

The organization of this paper is as follows: Section \ref{sec:system_model} presents the system model for MU-MIMO networks and describes the optimization problems. A DL-based precoder optimization approach is proposed in Section \ref{sec:DNN_structure_PCSI}, and it is extended to an end-to-end communication system in Section \ref{sec:ICSI}. Section \ref{sec:extension} provides some variations of the proposed approach for practical issues. Numerical results prove the effectiveness of proposed DL framework in Section \ref{sec:numerical_results}. Finally, the paper is concluded in Section \ref{sec:conclusion}.

\textit{Notations}: We employ uppercase boldface letters, lowercase boldface letters, and normal letters for matrices, column vectors, and scalar quantities, respectively. Complex-valued and real-valued matrices of size $C_1$-by-$C_2$ are denoted by $\mathbb{C}^{C_1 \times C_2}$ and $\mathbb{R}^{C_1 \times C_2}$, respectively. Also, $\|\mathbf{A}\|_F$ stands for the Frobenius norm of a matrix $\mathbf{A}$, and $(\cdot)^T$ and $(\cdot)^H$ respectively account for transpose and Hermitian operations.

\section{System Model} \label{sec:system_model}

We consider an FDD MU-MIMO downlink system where a BS equipped with $N_t$ antennas serves $K$ users each having $N_r$ antennas. We aim at optimizing the channel acquisition processes and downlink precoding strategies. We adopt limited feedback techniques \cite{Love:08,     Jindal:06, Ravindran:08} to obtain downlink CSI.
The BS first sends pilot sequences during a time interval $T_p$. Let $\mathbf{p}_{t_p}\in\mathbb{C}^{N_{t}\times1}$ be the pilot symbol vector at the $t_p$-th time instant $(t_p=1,\cdots,T_p)$. Denoting $\mathbf{H}_{k} \in \mathbb{C}^{N_r \times N_t}$ $(k=1,\cdots,K)$ as the channel matrix from the BS to user $k$, the received pilot signal at the $t_p$-th time instant $\mathbf{y}^{p}_{k,t_p} \in \mathbb{C}^{N_r \times 1}$ of the $k$-th user is given as \cite{Hassibi:03}
\begin{align}
\mathbf{y}_{k,t_p}^{p}=\mathbf{H}_k\mathbf{p}_{t_p}+\mathbf{n}_{k,t_p}^p, \label{eq:Y_k_train}
\end{align}
where $\mathbf{n}_{k,t_p}^p \in \mathbb{C}^{N_r \times 1}$ indicates the additive noise vector whose elements follow an independent Gaussian distribution with zero mean and variance $\sigma_{k}^{2}$. Defining the pilot matrix as $\mathbf{P}\triangleq[\mathbf{p}_1,\cdots,\mathbf{p}_{T_p}] \in \mathbb{C}^{N_t \times T_p}$, $\mathbf{P}$ is subject to the transmit power budget $E_{p}$ as
\begin{align}
\frac{1}{T_p}\sum_{t_p=1}^{T_p}\|\mathbf{p}_{t_p}\|^{2}= \frac{1}{T_p}\text{Tr}(\mathbf{P}\mathbf{P}^{H})=E_{p}. \label{eq:Ep_constraint}
\end{align}

Upon receiving pilot signals, the $k$-th user generates the feedback information $i_{k}\in\mathcal{B}\triangleq\{1,\cdots,2^{B}\}$, where $B$ denotes the number of feedback bits, and the user sends it back to the BS through finite-capacity uplink feedback channels. Stacking the received signals into a matrix form as $\mathbf{Y}_{k}^{p}\triangleq[\mathbf{y}^p_{k,1},\cdots,\mathbf{y}^p_{k,T_p}] \in \mathbb{C}^{N_r \times T_p}$ in \eqref{eq:Y_k_train}, we characterize the feedback information generation at the $k$-th user as a mapping $f_{\text{u},k}:\mathbb{C}^{N_{r}\times T_p}\rightarrow\mathcal{B}$, which is written by
\begin{align}
i_{k}=f_{\text{u},k}(\mathbf{Y}_{k}^{p})\in\mathcal{B}.\label{eq:fkuser}
\end{align}
In what follows, we will refer to $f_{\text{u},k}(\cdot)$ as the user operator.

Then, the BS collects all feedback information $\{i_{k},\forall k\}$ which encapsulates the channel information for all users to optimize the precoder. Assuming that $N_{t}\geq N_{r}$, the linear precoding matrix $\mathbf{V}_{k}\in\mathbb{C}^{N_{t}\times N_{r}}$ is employed to convey the data symbol vector $\mathbf{s}_{k}\sim\mathcal{CN}(\mathbf{0},\mathbf{I}_{N_{r}})\in\mathbb{C}^{N_{r}\times 1}$ for the $k$-th user. The precoding process at the BS is modeled as a mapping $f_{\text{BS}}:\mathcal{B}^{K}\rightarrow\mathbb{C}^{N_{t}\times KN_{r}}$~as
\begin{align}
\mathbf{V}=f_{\text{BS}}(\{i_{k},\forall k\}),\label{eq:fbs}
\end{align}
where $\mathbf{V}\triangleq[\mathbf{V}_1,\cdots,\mathbf{V}_K]\in \mathbb{C}^{N_t \times KN_r}$ accounts for the combined precoding matrix satisfying the transmit power constraint $E_{s}$ as
\begin{align}
\text{Tr}(\mathbf{V}\mathbf{V}^{H})=E_{s}.\label{eq:Es_constraint}
\end{align}

The precoded signal at the BS denoted by $\sum_{i=1}^{K}\mathbf{V}_{i}\mathbf{s}_{i}$ is broadcasted to the users through the downlink channels $\mathbf{H}$. Then, the received data signal $\mathbf{y}_{k}^{d}\in\mathbb{C}^{N_{r}\times1}$ at the $k$-th user is expressed~as
\begin{align}
\mathbf{y}_k^{d} = \mathbf{H}_k \mathbf{V}_{k}\mathbf{s}_{k} + \mathbf{H}_{k}\sum_{i=1,i\neq k}^{K}\mathbf{V}_{i}\mathbf{s}_{i} + \mathbf{n}_k, \label{eq:y_k_data}
\end{align}

where $\mathbf{n}_k \sim \mathcal{CN}(\mathbf{0}, \sigma_{k}^{2} \mathbf{I}_{N_r}) \in \mathbb{C}^{N_r \times 1} $ denotes the additive Gaussian noise. For given $\mathbf{H}_{k}$ and $\mathbf{V}$, the achievable data rate $R_{k}(\mathbf{H}_{k},\mathbf{V})$ at the $k$-th user is obtained by
\begin{align}
R_k(\mathbf{H}_{k},\mathbf{V}) \triangleq \log\det\left(\mathbf{I}_{N_r} + \mathbf{V}_k^{H}\mathbf{H}_k^{H}\mathbf{H}_k\mathbf{V}_k\left(\sigma_k^2\mathbf{I}_{N_r}+\displaystyle\sum_{i=1,i \neq k}^{K}\mathbf{H}_k\mathbf{V}_i\mathbf{V}_i^{H}\mathbf{H}_k^{H}\right)^{-1}\right). \label{eq:rate}
\end{align}

\subsection{Problem Description}
In this paper, we aim at maximizing the sum-rate performance $\sum_{k=1}^K R_k(\mathbf{H}_{k},\mathbf{V})$ by jointly optimizing the pilot matrix $\mathbf{P}$, the user operator $f_{\text{u},k}(\cdot)$ in \eqref{eq:fkuser}, and the BS precoding operator $f_{\text{BS}}(\cdot)$ in \eqref{eq:fbs}. The corresponding optimization task can be formulated as
\begin{subequations}
\begin{align}
\max_{\mathbf{P},\{f_{\text{u},k}(\cdot),\forall k\},f_{\text{BS}}(\cdot)}\ \ &\displaystyle\sum_{k=1}^K R_k(\mathbf{H}_{k},\mathbf{V}) \\
\text{subject to}\ &\ \eqref{eq:Ep_constraint},\eqref{eq:fkuser},\eqref{eq:fbs},\eqref{eq:Es_constraint}.
\end{align}\label{eq:sum-rate}
\end{subequations}

The nonconvexity of the sum-rate objective and the absence of closed-form expressions of the mappings $f_{\text{u},k}(\cdot)$ and $f_{\text{BS}}(\cdot)$ pose fundamental challenges in developing model-based optimization algorithms. As a consequence, existing studies have been confined to separate designs of each operator. To address this issue, we propose a DL-based approach which tackles an end-to-end optimization task including the pilots $\mathbf{P}$, the channel acquisition process $f_{\text{u},k}(\cdot)$, and the precoding optimizer $f_{\text{BS}}(\cdot)$. 
According to the CSIT available at the BS, we present problem \eqref{eq:sum-rate} in two different scenarios. First, in the ideal perfect CSIT case, we focus on designing the BS operator $f_{\text{BS}}(\cdot)$. Second, it is extended to a practical imperfect CSIT case that invokes joint optimization of the pilot and the two operators.

\subsection{Conventional Approaches for FDD MU-MIMO Systems} \label{sec:conventional_sys}
Before solving \eqref{eq:sum-rate}, we first review conventional approaches for FDD MU-MIMO systems, to provide useful insight into developing the proposed DL framework. First of all, each user estimates its corresponding channel matrix using the pilot signal \eqref{eq:Y_k_train} based on standard estimation techniques \cite{Hassibi:03, Tropp:07}. The resulting estimation $\hat{\mathbf{H}}_{k}$ is then quantized according to a predefined codebook $\mathcal{C}$ containing $2^B$ candidates of $\mathbf{H}_{k}$. The $k$-th user selects its feedback information $i_{k}$ as the index of an element in $\mathcal{C}$ that shows the best match with the estimated CSI, e.g., minimizing the chordal distance to $\hat{\mathbf{H}}_{k}$ \cite{Love:05}.

The BS then retrieves the CSI from feedback information. To maximize the sum-rate, $f_{\text{BS}}(\cdot)$ can be designed based on the WMMSE algorithm \cite{Christensen:08,Shi:11}. For notational simplicity, we explain the WMMSE algorithm assuming perfect CSIT $\mathbf{H}$. For a given weight $\mathbf{W}_k$, the sum-weighted mean square error (MSE) problem is formulated as
\begin{subequations}
\begin{align}
\min_{\mathbf{V}}\ \ &\displaystyle\sum_{k=1}^K \text{Tr}(\mathbf{W}_k\mathbf{E}_k(\mathbf{H}_k,\mathbf{V})) \label{eq:sum_WMSE} \\
\text{subject to }\ \ &\text{Tr}(\mathbf{V} \mathbf{V}^H) = E_s \label{eq:Es_constraint_wmmse},
\end{align}\label{eq:WMMSE}
\end{subequations}
where $\mathbf{E}_{k}(\mathbf{H}_k,\mathbf{V})$ stands for the mean square error matrix defined as
\begin{align}
\mathbf{E}_k(\mathbf{H}_k,\mathbf{V})=\left(\mathbf{I}_{N_r} + \mathbf{V}_{k}^{H}\mathbf{H}^{H}_{k}\left(\sigma^2_k\mathbf{I}_{N_r} + \displaystyle\sum_{i=1, i\neq k}^K \mathbf{H}_k\mathbf{V}_{i}\mathbf{V}_{i}^{H}\mathbf{H}_{k}^{H}\right)^{-1}\mathbf{H}_{k}\mathbf{V}_{k}\right)^{-1}.  \label{eq:mse}
\end{align}
The equivalence between \eqref{eq:sum-rate} and \eqref{eq:WMMSE} can be established by setting $\mathbf{W}_{k}=\mathbf{E}_{k}(\mathbf{H}_k,\mathbf{V})^{-1}$.

The WMMSE algorithm adopts an alternating optimization procedure between $\mathbf{W}_{k}$ and $\mathbf{V}_{k}$. At each iteration, the receive filter $\mathbf{U}_k$, the weight matrix $\mathbf{W}_k$, and the precoding matrix $\mathbf{V}_{k}$ are updated as
\begin{align}
\mathbf{U}_k&=\mathbf{V}_{k}^{H}\mathbf{H}_{k}^{H} \left(\sigma^2_k\mathbf{I}_{N_r} + \displaystyle\sum_{i=1}^K \mathbf{H}_k\mathbf{V}_{i}\mathbf{V}_{i}^{H}\mathbf{H}_{k}^{H}\right)^{-1},\label{eq:receive_filter} \\
\mathbf{W}_k&= \mathbf{E}_k(\mathbf{H}_k,\mathbf{V})^{-1}, \label{eq:eq_weight}\\
\mathbf{V}_{k}&=\gamma\left(\displaystyle\sum_{k=1}^{K}\textbf{H}_{k}^H\textbf{U}_{k}^H\mathbf{W}_{k}\textbf{U}_{k}\textbf{H}_{k}+\beta \mathbf{I}_{N_t}\right)^{-1}\textbf{H}_{k}^{H}\mathbf{U}_{k}^{H}\textbf{W}_{k} \nonumber \\ &\triangleq\gamma \tilde{\mathbf{V}}_k, \label{eq:wmmse_precoder1}
\end{align}
where $\gamma \triangleq \sqrt{E_s/\text{Tr}(\tilde{\mathbf{V}}\tilde{\mathbf{V}}^H)}$ represents a constant for equal power constraint of \eqref{eq:Es_constraint_wmmse} and $\beta = \sum_{k=1}^{K} \frac{\sigma^2_k}{E_s}\text{Tr}(\textbf{W}_{k}\textbf{U}_{k}\textbf{U}_{k}^H)$ is a regularization factor. The above iteration converges to a locally optimum point of the sum-rate maximization problem \eqref{eq:sum-rate}. For the case of perfect CSIT, the WMMSE algorithm can serve as the performance benchmark as it maximizes the sum-rate. However, the computational complexity is generally high, especially in high SNR. Another critical issue of the WMMSE algorithm is that the performance significantly deteriorates for imperfect CSIT.

\subsection{DNN Preliminaries}
We briefly introduce the notation for fully-connected (FC) DNNs. Let $\mathbf{Z}=\mathcal{F}(\mathbf{X};\mathbf{\theta})$ denote a DNN which processes an input matrix $\mathbf{X} \in \mathbb{C}^{X_1 \times X_2}$ with a trainable parameter set $\mathbf{\theta}$ and yields an output matrix $\mathbf{Z} \in \mathbb{C}^{Z_1 \times Z_2}$. The input matrix $\mathbf{X}$ is first vectorized into a real-value representation as $\mathbf{x}=\text{vec}([\Re\{\mathbf{X}\}^T, \Im\{\mathbf{X}\}^T]^{T}) \in \mathbb{R}^{2X_1 X_2}$ and the vectorized real-value prepresentation of the output $\mathbf{z} = \text{vec}([\Re\{\mathbf{Z}\}^T,\Im\{\mathbf{Z}\}^T]^{T}) \in \mathbb{R}^{2Z_1Z_2}$ is computed by the DNN, where $\Re\{\cdot\}$ and $\Im\{\cdot\}$ stand for the real and imaginary parts, respectively. The computations of an $L$-layer DNN can be written by
\begin{align}
\mathbf{z} = a_{L}(\mathbf{\Phi}_{L}( \cdots  (\mathbf{\Phi}_{2}a_1(\mathbf{\Phi}_{1}\mathbf{x}+\mathbf{b}_1)+ \mathbf{b}_2) \cdots )+ \mathbf{b}_{L}),\label{eq:fnn}
\end{align}
where $a_{l}(\cdot)$ for $l=1,\cdots,L$ indicates an activation function at the $l$-th layer with dimension $D_l$, and $\mathbf{\Phi}_l\in \mathbb{R}^{D_{l} \times D_{l-1}}$ and $\mathbf{b}_l \in \mathbb{R}^{D_l}$ respectively represent a weight matrix and a bias vector, which form a set of trainable variables $\mathbf{\mathbf{\theta}}\triangleq\{\mathbf{\Phi}_l, \mathbf{b}_l , \forall l \}$. The final output matrix $\mathbf{Z}$ is then readily obtained from its vector representation $\mathbf{z}$.

\section{Precoder Optimization} \label{sec:DNN_structure_PCSI}
In this section, we first consider the ideal scenario where the perfect CSIT $\mathbf{H}$ is available at the BS, which means that the pilot matrix $\mathbf{P}$ and the user operators $\{f_{\text{u},k}(\cdot),\forall k\}$ in \eqref{eq:sum-rate} are assumed to be perfect. We focus on the precoder optimization, i.e., the BS operator $f_{\text{BS}}(\cdot)$. A DNN module at the BS denoted by $\mathcal{G}_{\text{BS}}(\cdot)$ is introduced to replace the mapping $f_{\text{BS}}(\cdot)$ with neural calculations. One possible solution for constructing the BS DNN $\mathcal{G}_{\text{BS}}(\cdot)$ is to apply a naive FC DNN that directly produces the precoding matrix $\mathbf{V}$. This approach, however, turns out to be ineffective even in a simple single-antenna user case \cite{Kim:20}. To address this issue, we carefully design the architecture and training strategies of the BS DNN, inspired from the mechanism of the WMMSE algorithm \cite{Christensen:08}. In what follows, we first describe the structure of the BS DNN, along with the training algorithm.

\begin{figure}
\begin{center}
\includegraphics[width=4in]{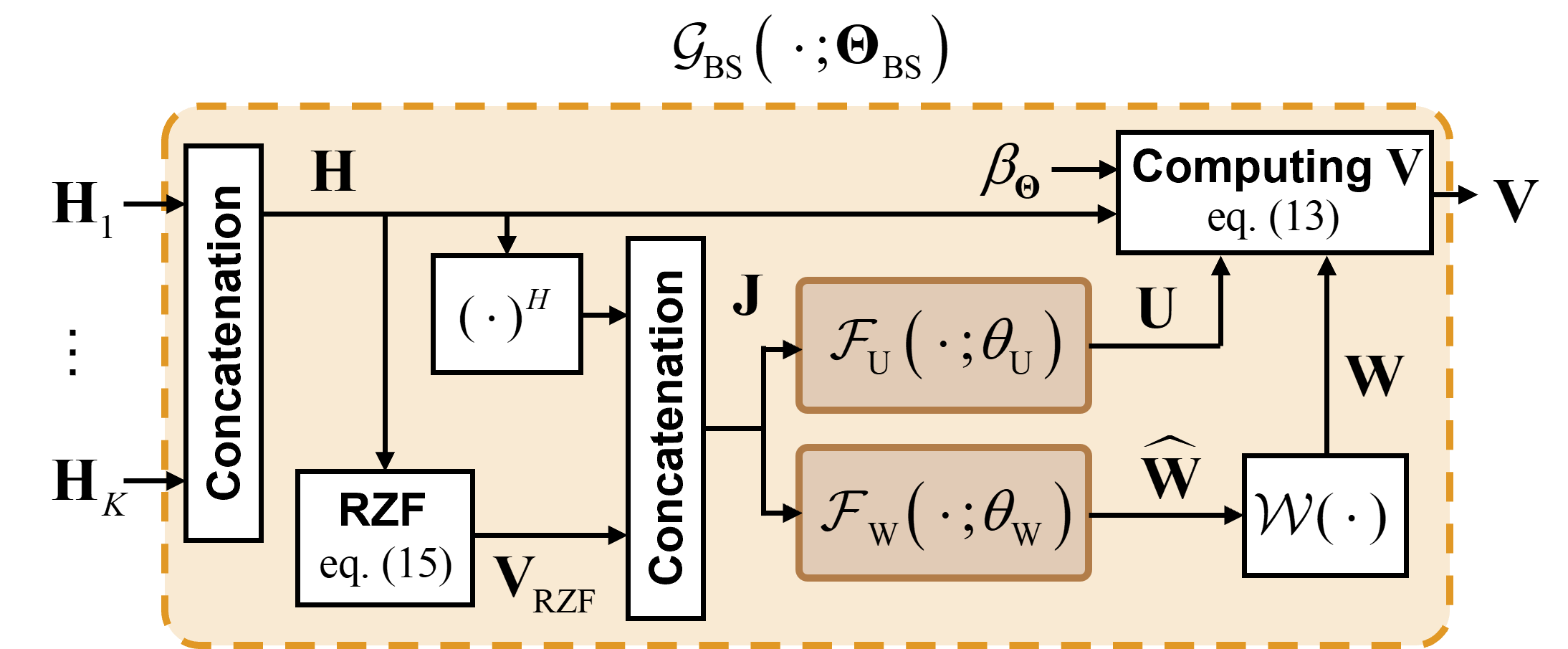}
\end{center}
\caption{DNN structure for precoder generation in MU-MIMO systems.}
\label{figure:DNN_Precoder}
\end{figure}

\subsection{DNN Structure}
Fig. \ref{figure:DNN_Precoder} illustrates the proposed architecture of the BS DNN which yields the precoding matrix $\mathbf{V}$ for a given CSI $\mathbf{H}=[\mathbf{H}_1^T,\cdots,\mathbf{H}_K^T]^T \in \mathbb{C}^{KN_r \times N_t}$. Instead of determining the precoding matrix $\mathbf{V}$ directly, the proposed learning structure exploits the expert knowledge of the WMMSE algorithm. We obtain the weight matrix $\mathbf{W}_{k}$ \eqref{eq:eq_weight} and the receive filter $\mathbf{U}_{k}$ \eqref{eq:receive_filter} to form the precoding matrix $\mathbf{V}_{k}$ \eqref{eq:wmmse_precoder1}. The BS DNN first utilizes the regularized zero forcing (RZF) precoder $\mathbf{V}_{\text{RZF}}$ \cite{Peel:05} from the input CSI $\mathbf{H}$ as
\begin{align}
\mathbf{V}_{\text{RZF}} = \gamma_{\text{RZF}}\mathbf{H}^H \left( \mathbf{H}\mathbf{H}^H + \beta_{\text{RZF}} \mathbf{I}_{KN_r}\right)^{-1}, \label{eq:RZF}
\end{align}
where $\gamma_{\text{RZF}}$ stands for the normalization constant for satisfying the transmit power constraint $\text{Tr}(\mathbf{V}_{\text{RZF}}\mathbf{V}_{\text{RZF}}^{H})=E_{s}$ in \eqref{eq:Es_constraint} and $\beta_{\text{RZF}}=\sum_{k=1}^{K}\frac{\sigma_{k}^{2}N_{r}}{E_{s}}$ denotes the regularization factor. Due to simplicity, zero forcing (ZF) has been adopted as initialization schemes for other optimization algorithms \cite{Xia:19} and as input features for DNNs \cite{Hassan:11}. Motivated by these results, we exploit $\mathbf{V}_{\text{RZF}}$ as side information to the BS DNN along with the CSI $\mathbf{H}$. Such extra side information helps the BS DNN effectively extract crucial features of the optimal precoding matrices. The effectiveness of the RZF input will be numerically verified in Sec. \ref{sec:numerical_results}.

In the proposed BS DNN, we pass $\mathbf{J}=\left[ \mathbf{H}^{H}, \mathbf{V}_{\text{RZF}} \right] \in \mathbb{C}^{N_t \times 2KN_r}$ to two FC DNNs $\mathcal{F}_{\text{W}}(\cdot;\mathbf{\theta}_{\text{W}})$ and $\mathcal{F}_\text{U}(\cdot;\mathbf{\theta}_{\text{U}})$, which are respectively dedicated to the computations of the weight matrix $\mathbf{W}_{k}$ in \eqref{eq:eq_weight} and the receive filter $\mathbf{U}_{k}$ in \eqref{eq:receive_filter}. For the first DNN $\mathcal{F}_\text{W}(\cdot;\mathbf{\theta}_{\text{W}})$ with trainable parameter set $\mathbf{\theta}_{\text{W}}$ having $L_{\text{W}}$ hidden layers, its output $\hat{\mathbf{W}}=[ \hat{\mathbf{W}}_1,\cdots,\hat{\mathbf{W}}_K] \in \mathbb{C}^{N_r \times KN_r}$ is obtained as
\begin{align}
\hat{\mathbf{W}}=\mathcal{F}_\text{W}(\mathbf{J};\mathbf{\theta}_{\text{W}}).\label{eq:FW}
\end{align}
Since $\mathbf{W}_{k}$ of the WMMSE algorithm \eqref{eq:eq_weight} is Hermitian, we construct the weight matrix of the proposed DNN method as
\begin{align}
\mathbf{W}_k = \hat{\mathbf{W}}_k \hat{\mathbf{W}}_k^H + \mathbf{I}_{N_r}\triangleq \mathcal{W}(\hat{\mathbf{W}}_k),\label{eq:Wtilde}
\end{align}
where an identity matrix $\mathbf{I}_{N_r}$ is added to follow the structure of the WMMSE algorithm \eqref{eq:mse}.

The second DNN $\mathcal{F}_\text{U}(\cdot;\mathbf{\theta}_{\text{U}})$, which has $L_{\text{U}}$ hidden layers and the trainable parameter set $\mathbf{\theta}_{\text{U}}$, calculates the receive filter $\mathbf{U}=[\mathbf{U}_1,\cdots,\mathbf{U}_K ] \in \mathbb{C}^{N_r \times KN_r}$ as
\begin{align}
\mathbf{U}=\mathcal{F}_\text{U}(\mathbf{J};\mathbf{\theta}_{\text{U}}).\label{eq:Utilde}
\end{align}
We then compute the precoding matrix $\mathbf{V}$ based on the WMMSE algorithm \eqref{eq:wmmse_precoder1} as \eqref{eq:DNN_WMMSE},

\begin{align}
\mathbf{V}&=\gamma_{\mathbf{\mathbf{\Theta}}}\left( \displaystyle \sum^{K}_{k=1}\mathbf{H}_k^H\mathbf{U}_k^H\mathbf{W}_k \mathbf{U}_k\mathbf{H}_k + (\beta+\beta_{\mathbf{\mathbf{\Theta}}}^{2})\mathbf{I}_{N_t}\right)^{-1} \left[\mathbf{H}_1^H \mathbf{U}_1^H \mathbf{W}_1, \cdots, \mathbf{H}_K^H \mathbf{U}_K^H \mathbf{W}_K \right] \label{eq:DNN_WMMSE}\\
&\triangleq\gamma_{\mathbf{\Theta}} \tilde{\mathbf{V}} \nonumber
\end{align}

where $\gamma_{\mathbf{\Theta}}\triangleq\sqrt{E_s/\text{Tr}(\tilde{\mathbf{V}}\tilde{\mathbf{V}}^H)}$ denotes the power normalization constant such that $\text{Tr}(\mathbf{V}\mathbf{V}^{H})=E_{s}$ and $\beta_{\mathbf{\Theta}}\in\mathbb{R}$ represents a trainable variable that regulates possible channel uncertainties, which will be described later. Since the input feature $\mathbf{J}$ of $\mathcal{F}_\text{W}(\cdot;\mathbf{\theta}_{\text{W}})$ and $\mathcal{F}_\text{U}(\cdot;\mathbf{\theta}_{\text{U}})$ can be determined from the CSI feature $\mathbf{H}$, the BS DNN $\mathcal{G}_{\text{BS}}(\cdot)$ can be modeled as a mapping from $\mathbf{H}$ to $\mathbf{V}$ with trainable parameters $\mathbf{\mathbf{\Theta}}_{\text{BS}}\triangleq\{\mathbf{\theta}_{\text{W}},\mathbf{\theta}_{\text{U}},\beta_{\mathbf{\mathbf{\Theta}}}\}$. Then, the corresponding input-output relationship can be written by
\begin{align}
\mathbf{V} \triangleq \mathcal{G}_{\text{BS}}(\mathbf{H};\mathbf{\mathbf{\Theta}}_{\text{BS}}). \label{eq:mapping_precoderNN}
\end{align}

\subsection{Training Strategy} \label{sec:training_PCSI}
The BS DNN is trained to maximize the sum-rate performance, or equivalently, to minimize the sum-weighted MSE \eqref{eq:WMMSE}. By substituting \eqref{eq:mapping_precoderNN} into \eqref{eq:sum_WMSE}, the training problem for given weight matrices $\mathbf{W}_{\mathbf{\mathbf{\Theta}},k}$ can be formulated as
\begin{align}
\min_{\mathbf{\mathbf{\Theta}}_{\text{BS}}} &\ \ \displaystyle \sum_{k=1}^{K} \text{Tr}\left( \mathbf{W}_{\mathbf{\mathbf{\Theta}}, k}\mathbf{E}_{k}\Big(\mathbf{H}_k, \mathcal{G}_{\text{BS}}(\mathbf{H};\mathbf{\mathbf{\Theta}}_{\text{BS}})\Big)\right) \label{eq:DNN_sum_rate},
\end{align}
where the transmit power constraint \eqref{eq:Es_constraint_wmmse} is removed due to the normalization factor $\gamma_{\mathbf{\mathbf{\Theta}}}$ in \eqref{eq:DNN_WMMSE}.

\begin{figure}
\begin{center}
\includegraphics[width=2.7in]{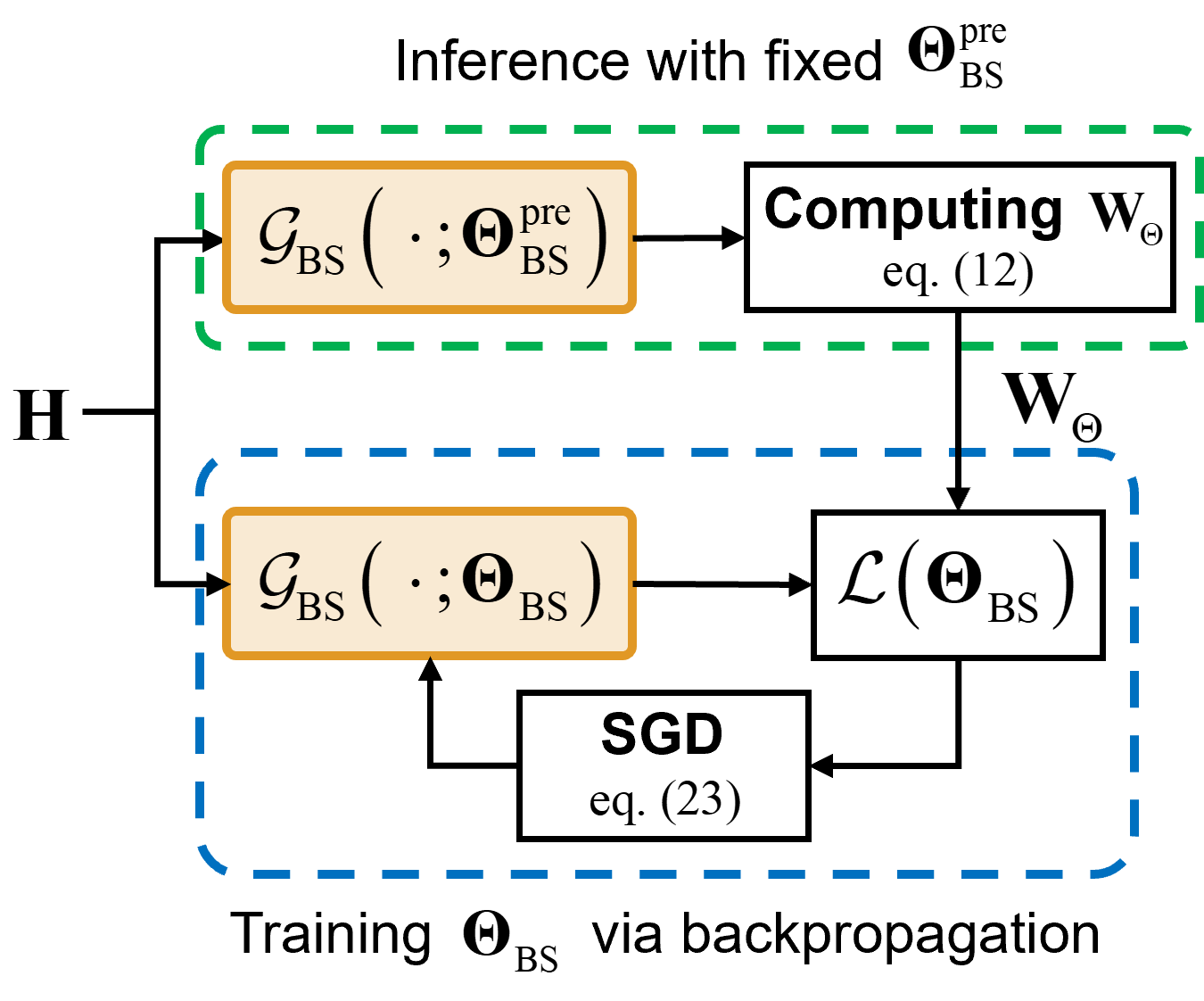}
\end{center}
\caption{Schematic diagram of BS DNN training at each stage.}
\label{figure:DNN_training}
\end{figure}

To bridge \eqref{eq:DNN_sum_rate} to the sum-rate maximization problem, we need to include the recursive relationship $\mathbf{W}_{\mathbf{\mathbf{\Theta}}, k}=\mathbf{E}_{k}\big(\mathbf{H}_k,\mathcal{G}_{\text{BS}}(\mathbf{H};\mathbf{\mathbf{\Theta}}_{\text{BS}})\big)^{-1}$ in the training of the BS DNN. We employ the concept of the multi-stage training strategy \cite{Barshan:15} in which the trained DNN at the previous stage, e.g., the trained parameters and the training loss function, is utilized for the initialization of the DNN at the current training stage. Such multiple training stages accumulate the knowledge acquired in the previous DNNs and successfully improve the training performance. In our case, we transfer the weight matrix $\mathbf{W}_{\mathbf{\mathbf{\Theta}}, k}$ obtained by the BS DNN at the previous stage to the loss function at the current stage so that it can be exploited for constructing the sum-weighted MSE function \eqref{eq:DNN_sum_rate}. This can be viewed as the alternating optimization strategy of the WMMSE algorithm \eqref{eq:receive_filter}--\eqref{eq:wmmse_precoder1} with gradual refinements of the optimization objective.

To this end, we employ two BS DNNs: one for obtaining $\mathbf{W}_{\mathbf{\Theta}}=[\mathbf{W}_{\mathbf{\Theta},1},\cdots,\mathbf{W}_{\mathbf{\Theta},K}]$ and the other for training as illustrated in Fig. \ref{figure:DNN_training}. To avoid notational confusion, we denote the parameter set of the BS DNN for determining $\mathbf{W}_{\mathbf{\Theta}}$ as $\mathbf{\Theta}_{\text{BS}}^{\text{pre}}$. At the beginning of each training stage, $\mathbf{\Theta}_{\text{BS}}$ trained at the previous stage is copied to $\mathbf{\Theta}^{\text{pre}}_{\text{BS}}$, which is fixed to produce the weight $\mathbf{W}_{\mathbf{\Theta}}$. Then, the weight $\mathbf{W}_{\mathbf{\Theta}, k}$ is computed from \eqref{eq:eq_weight} as $\mathbf{W}_{\mathbf{\Theta}, k}=\mathbf{E}_{k}\big(\mathbf{H}_k,\mathcal{G}_{\text{BS}}(\mathbf{H};\mathbf{\Theta}_{\text{BS}}^{\text{pre}})\big)^{-1}$, and the corresponding loss function $\mathcal{L}(\mathbf{\Theta}_{\text{BS}})$ is set to the expected sum-weighted MSE averaged over the channel distribution $\mathbf{H}$ given by
\begin{align}
\mathcal{L}(\mathbf{\Theta}_{\text{BS}})\triangleq\mathbb{E}_{\mathbf{H}}\left[ \displaystyle\sum_{k=1}^K\text{Tr} \Bigg( \mathbf{W}_{\mathbf{\Theta}, k}
\mathbf{E}_{k}\Big(\mathbf{H}_k, \mathcal{G}_{\text{BS}}(\mathbf{H};\mathbf{\Theta}_{\text{BS}})\Big) \Bigg) \right], \label{eq:DNN_sum-weighted MSE}
\end{align}
where $\mathbb{E}_{X}[\cdot]$ indicates the expectation operator over a random variable $X$. In \eqref{eq:DNN_sum-weighted MSE}, $\mathbf{W}_{\mathbf{\Theta}, k}$ of the previous stage is utilized so that the optimality condition \eqref{eq:eq_weight} can be gradually established by the precoding matrix $\mathbf{V}=\mathcal{G}_{\text{BS}}(\mathbf{H};\mathbf{\Theta}_{\text{BS}})$ as the number of training stages increases. Thus, the alternating optimization processes in \eqref{eq:receive_filter}--\eqref{eq:wmmse_precoder1} of the WMMSE algorithm, which recursively updates $\mathbf{W}$ and $\mathbf{V}$ by fixing one at a time, is converted to the BS DNN training. To be specific, the training objective \eqref{eq:DNN_sum-weighted MSE} can be viewed as the precoder update step of the WMMSE method for a fixed weight matrix $\mathbf{W}_{\boldsymbol{\Theta}}$ with the DNN parameter $\boldsymbol{\Theta}_{\text{BS}}^{\text{pre}}$ optimized at the previous stage. Then, we recover the optimum weight matrix with a given precoding $\mathbf{V}=\mathcal{G}_{\text{BS}}(\mathbf{H};\boldsymbol{\Theta}_{\text{BS}})$ from \eqref{eq:eq_weight}. Such an alternating optimization process is repeated over multiple training stages. The convergence of this multi-stage training algorithm is guaranteed by the conventional block coordinate descent (BCD) framework \cite{Tseng:01}.

The BS DNN is trained by standard gradient-based optimization, e.g., the mini-batch stochastic gradient descent (SGD) method. At each epoch of the training stages, the BS DNN parameter in $\mathbf{\Theta}_{\text{BS}}$ is updated as
\begin{align}
\mathbf{\Theta}_{\text{BS}} &\leftarrow \mathbf{\Theta}_{\text{BS}} -\eta \nabla_{\mathbf{\Theta}_{\text{BS}}} \mathbb{E}_{\mathbf{H}}\left[ \displaystyle\sum_{k=1}^K\text{Tr} \Bigg( \mathbf{W}_{\mathbf{\Theta}, k}
\mathbf{E}_{k}\Big(\mathbf{H}_k, \mathcal{G}_{\text{BS}}(\mathbf{H};\mathbf{\Theta}_{\text{BS}})\Big) \Bigg) \right], \label{eq:SGD_mth_Tr}
\end{align}
where $\eta$ denotes the learning rate. The parameter update \eqref{eq:SGD_mth_Tr} is iterated until the loss $\mathcal{L}(\mathbf{\Theta}_{\text{BS}})$ converges. Then, the next stage is conducted with the trained parameter set $\mathbf{\Theta}_{\text{BS}}$, and this procedure is repeated until convergence. We summarize our proposed training strategy in Algorithm \ref{algo:BS_NN_fine}. The training data can be collected by generating channel matrices which follow the probability density functions of any given channel models or experimentally measuring channel samples.

\begin{algorithm} [bt!]
\caption{Multi-stage training policy} \label{algo:BS_NN_fine}
\begin{algorithmic}[1]
\STATE {} Initialize $\mathbf{\Theta}_{\text{BS}}=\mathbf{\Theta}_{\text{BS}}^{\text{init}}$.
\STATE {} \textbf{Repeat}
\STATE {} ~~ Set $\mathbf{\Theta}_{\text{BS}}^{\text{pre}}=\mathbf{\Theta}_{\text{BS}}$.
\STATE {} ~~ \textbf{Repeat}
\STATE {} ~~~~ Sample the mini-batch set $\mathcal{H}_{t}$ from the training dataset.
\STATE {} ~~~~ Compute $\mathbf{W}_{\mathbf{\Theta}, k}=\mathbf{E}_{k}\big(\mathbf{H}_k,\mathcal{G}_{\text{BS}}(\mathbf{H};\mathbf{\Theta}_{\text{BS}}^{\text{pre}})\big)^{-1}$, $\forall k, \forall \mathbf{H}\in\mathcal{H}_{t}$.
\STATE {} ~~~~ Update $\mathbf{\Theta}_{\text{BS}}$ from \eqref{eq:SGD_mth_Tr}.
\STATE {} ~~ \textbf{Until} convergence
\STATE {} \textbf{Until} convergence
\end{algorithmic}
\end{algorithm}

To perform \eqref{eq:SGD_mth_Tr} at the first stage, we need an appropriate initialization of $\mathbf{\Theta}_{\text{BS}}$. This is achieved by determining the initial BS DNN parameter set $\mathbf{\Theta}_{\text{BS}}^{\text{init}}$ with $\mathbf{W}_{\mathbf{\Theta}, k}=\mathbf{I}_{N_r}$, leading to the sum-MSE minimization. The parameter set $\mathbf{\theta}_W$, which is dedicated to the identification of $\mathbf{W}_{k}$ in \eqref{eq:Wtilde}, is not adjusted in this initial training step, since we can obtain $\mathbf{W}_{k}=\mathbf{I}_{N_r}$ without $\mathbf{\theta}_{\text{W}}$. Then, the multi-stage training is conducted for the sum-weighted MSE minimization. We progress the training stages until the average sum-rate $\mathbb{E}_{\mathbf{H}}\left[\sum^{K}_{k=1} R_k(\mathbf{H}_k,\mathbf{V})\right]$ converges. Then, the BS DNN parameter set $\mathbf{\Theta}_{\text{BS}}$ trained at the final stage is stored at the BS to conduct the real-time precoder calculation \eqref{eq:mapping_precoderNN}. Thus, the number of the training stages do not affect the computational complexity of the trained BS DNN.

\begin{remark}
In \cite{Hu:20}, the unfolding technique with cascaded DNN modules was employed to characterize existing iterative optimization algorithms, i.e., the WMMSE algorithm. An output of each module is modeled as an intermediate precoder solution of each iteration step, which becomes the input to the subsequent DNN modules. For this reason, all constituting DNN modules for the unfolding method need to be stored to obtain the final precoder output at the BS, which leads to increased computational complexity for online implementation. On the contrary, our proposed training strategy requires a single DNN module only, which results in computational complexity reductions compared to existing DL based solutions, and this will be validated in Sec. \ref{sec:numerical_results}.
\end{remark}

\section{End-to-End System Optimization} \label{sec:ICSI}
\begin{figure*}
\begin{center}
\includegraphics[width=6.2in]{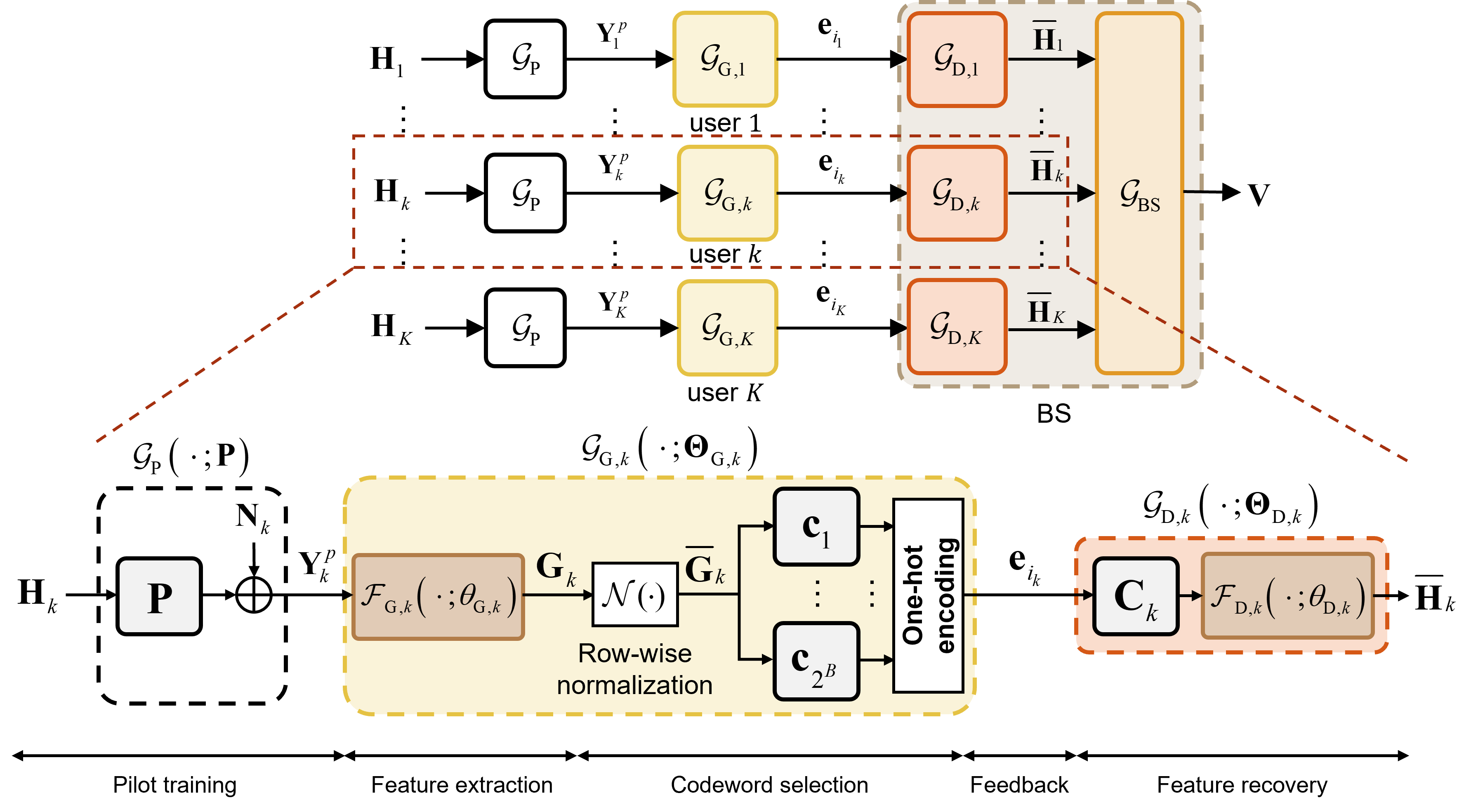}
\end{center}
\caption{Proposed DNN structure for practical FDD systems.}
\label{figure:DNN_Uplink_ICSI}
\end{figure*}

In the previous section, we have focused on the BS DNN design assuming the perfect CSIT case. In this section, we consider a general and practical scenario of imperfect CSIT where the pilot matrix, the user operator, and the BS operator need to be jointly designed. To this end, we propose a novel DNN structure as illustrated in Fig. \ref{figure:DNN_Uplink_ICSI}. Specifically, we additionally deploy the pilot NN, the user DNN $\{\mathcal{G}_{\text{G},k}(\cdot), \forall k\}$, and the dequantizer DNN $\{\mathcal{G}_{\text{D},k}(\cdot),\forall k\}$, which are placed before the BS DNN $\mathcal{G}_{\text{BS}}(\cdot)$ in \eqref{eq:mapping_precoderNN}. The details of these NN modules are explained in the following.

\subsection{Pilot Training}
The pilot NN obtains the pilot matrix $\mathbf{P}$, which is considered as a trainable variable. Then, \eqref{eq:Y_k_train} can be regarded as a forwardpass computation of $\mathbf{H}_{k}$ through a single-layer NN with the weight $\mathbf{P}$ and the randomized bias $\mathbf{N}_{k}$ \cite{Chun:19, Li:19}. To train the pilot matrix $\mathbf{P}$ using two NNs, we construct the pilot matrix with two NNs, of which weight matrices correspond to the real and imaginary parts of $\mathbf{P}$, respectively \cite{Li:19}. Thus, we can rewrite \eqref{eq:Y_k_train} in an NN formulation as
\begin{align}
\mathbf{Y}_{k}^{p}=\mathcal{G}_{\text{P}}(\mathbf{H}_{k};\mathbf{P}),\label{eq:pilotNN}
\end{align}
where $\mathbf{P}$ is subject to the transmit power constraint \eqref{eq:Ep_constraint}. The calculation of the pilot NN in \eqref{eq:pilotNN} depends on the channel matrix input, which is required only in the training step for determining the backward propagation through the trainable variable $\mathbf{P}$. Once the pilot NN is trained, the resulting $\mathbf{P}$ can be exploited as the pilot matrix in the test step without any prior knowledge of the downlink channels.

\subsection{Feature Extraction}
Upon receiving the pilot signal $\mathbf{Y}^{p}_k$, the $k$-th user generates the discrete information $i_{k}\in\mathcal{B}$ as in \eqref{eq:fkuser} for the $B$-bit limited feedback. Such a channel quantization procedure is carried out by the user operator $f_{\text{u},k}(\cdot)$, which is approximated by the user DNN $\mathcal{G}_{\text{G},k}(\cdot)$. It first extracts useful features of the received pilot signal $\mathbf{Y}^{p}_k$ by employing a FC DNN $\mathcal{F}_{\text{G},k}(\cdot;\mathbf{\theta}_{\text{G},k})$ with $L_{\text{G}}$ hidden layers. As will be discussed later, it is trained to predict the actual CSI $\mathbf{H}_{k}$ from the received pilot signal $\mathbf{Y}^{p}_k$. Therefore, the corresponding output $\mathbf{G}_k \in \mathbb{C}^{N_r \times N_t}$ acts as the estimated $\mathbf{H}_{k}$, which is given as
\begin{align}
\mathbf{G}_k=\mathcal{F}_{\text{G},k}(\mathbf{Y}_{k}^{p};\mathbf{\theta}_{\text{G},k})
=\mathcal{F}_{\text{G},k}(\mathcal{G}_{\text{P}}(\mathbf{H}_{k};\mathbf{P});\mathbf{\theta}_{\text{G},k}).\label{eq:Fenc}
\end{align}

\subsection{Codeword Selection in the Codebook}
The final output of the user DNN must be a discrete value $i_{k}\in\mathcal{B}$ for any given continuous-valued input $\mathbf{Y}_{k}^{p}$. One possible solution is a binarization technique that designs a neural quantizer yielding binary vectors for arbitrary continuous-valued inputs. Such a method has been widely employed in recent works on DL-based FDD systems \cite{Jang:20, Sohrabi:21, Kong:21}. This approach leads to efficient backpropagation algorithms for handling unavailable gradients of binarization operators such as the unit-step function. The resulting output becomes the quantized values of the input channel matrices, i.e., integer numbers or binary vectors, which can be utilized as an alternative representation of the selected codewords. These are further processed by DNNs to generate the target optimization variables, e.g., precoding matrices \cite{Jang:20, Sohrabi:21, Kong:21}. However, DNNs would not be suitable for handling binary-valued latent features, since they cannot represent any statistical properties of actual channel matrices. Furthermore, existing methods \cite{Jang:20, Sohrabi:21, Kong:21} are unable to provide any useful insights into practical limited feedback systems, e.g., optimized channel codebooks and selected codewords.

To overcome such issues, we develop a novel DNN method, which designs more flexible codewords based on the vector quantization operations of conventional limited feedback systems \cite{Jindal:06, Ravindran:08}. Unlike the binarization techniques that output discrete information, the proposed vector quantization approach ensures that a finite-alphabet channel codebook is learned in a vector space. Elements of the learned codebook act as continuous-valued channel codewords that encode statistical properties of the channel dataset. This can be viewed as a generalization of word embedding techniques \cite{Mikolov:13} which express words as flexible vector-valued representations suitable for DNNs. As a consequent, degree of freedom for a system design can be remarkably improved compared to existing binarization methods. We introduce the channel codebook matrix $\mathbf{C}_{k}=[\mathbf{c}_{k,1}, \cdots, \mathbf{c}_{k,2^B}]\in \mathbb{C}^{N_t \times 2^B}$ as a trainable parameter, where each column $\mathbf{c}_{k,j}$ stands for a codeword with $\Vert \mathbf{c}_{k,j} \Vert^2 = 1$. To select a codeword, we first apply the row-wise normalization $\mathcal{N}(\cdot)$ to the CSI feature $\mathbf{G}_k$ and obtain $\bar{\mathbf{G}}_{k}=\mathcal{N}(\mathbf{G}_{k})$, so that each row corresponds to a codeword candidate. Then, a codeword is chosen according to the correlation measure between the normalized feature $\bar{\mathbf{G}}_{k}$ and the codeword $\{\mathbf{c}_{k,j},\forall j\}$. Thus, the feedback information $i_{k}$ is given as
\begin{align}
i_{k}=\argmax_{j \in \mathcal{B}} \left\|\bar{\mathbf{G}}_{k}\mathbf{c}_{k,j}\right\|^2.\label{eq:iksel}
\end{align}

The trained $\mathbf{C}_{k}$ can be readily utilized as the optimized codebook along with the codeword selection rule \eqref{eq:iksel}. However, the selection operation in \eqref{eq:iksel} nulls the gradients with respect to $\mathbf{C}_{k}$, thereby posing a gradient vanishing issue. To address this issue, we approximate \eqref{eq:iksel} as a vector $\mathbf{e}_{i_{k}} \in \mathbb{R}^{2^B}$ whose $l$-th element $e_{i_{k},l}$ is expressed as
\begin{align}
e_{i_{k},l} = \frac{\left\| \bar{\mathbf{G}}_{k} \mathbf{c}_{k,l} \right\|^{\alpha}}{ \sum_{j\in\mathcal{B}} \left\|\bar{\mathbf{G}}_k \mathbf{c}_{k,j} \right\|^{\alpha}}, \label{eq:elk}
\end{align}
where a nonnegative hyperparameter $\alpha$ controls the smoothness.

As $\alpha$ goes to infinity, $\mathbf{e}_{i_{k}}$ approaches a one-hot encoding of $i_{k}$, i.e., $\mathbf{e}_{i_{k}}$ becomes an all zero vector except the $i_{k}$-th element being replaced by one. The gradient of the approximation \eqref{eq:elk} is nonzero for arbitrary $\alpha>0$. Although a large $\alpha$ leads to an accurate approximation of \eqref{eq:iksel}, it would incur the gradient vanishing issue. Thus, $\alpha$ needs to be carefully chosen as a moderate number. Finally, the input-output mapping of the user DNN $\mathcal{G}_{\text{G},k}(\cdot)$ is specified by the DNN $\mathcal{F}_{\text{G},k}(\cdot;\mathbf{\theta}_{\text{G},k})$, the normalization $\mathcal{N}(\cdot)$, and the relaxed quantization \eqref{eq:elk}. Denoting $\mathbf{\Theta}_{\text{G},k}\triangleq\{\mathbf{\theta}_{\text{G},k},\mathbf{C}_{k}\}$ as the set of the trainable parameters of the user DNN, we have
\begin{align}
\mathbf{e}_{i_k}=\mathcal{G}_{\text{G},k}(\mathbf{Y}^{p}_k;\mathbf{\Theta}_{\text{G},k}). \label{eq:user_NN}
\end{align}

\subsection{Feature Recovery}
The BS DNN presented in Sec. \ref{sec:DNN_structure_PCSI} has been dedicated to the perfect CSIT case, and thus it could not handle the quantized CSI $\mathbf{e}_{i_{k}}$. Therefore, a careful refinement of the BS DNN is needed to adapt to the general imperfect CSIT scenario. To this end, we add dequantization steps in front of \eqref{eq:mapping_precoderNN} to generate continuous-valued CSI features suitable for the BS DNN. Such a task is carried out by the dequantizer DNN $\mathcal{G}_{\text{D},k}(\cdot)$ which executes the reverse operation of the user DNN $\mathcal{G}_{\text{G},k}(\cdot)$. We first determine the codeword $\mathbf{c}_{k,i_k}$ selected by the $k$-th user as 
\begin{align}
\mathbf{c}_{k,i_k}= \mathbf{C}_k \mathbf{e}_{i_k}. \label{eq:ckik}
\end{align}
The resulting codeword $\mathbf{c}_{k,i_k}$ is then exploited as the input of the FC DNN $\mathcal{F}_{\text{D},k}(\cdot)$ to obtain the CSI feature $\bar{\mathbf{H}}_k=\mathcal{F}_{\text{D},k}(\mathbf{c}_{k,i_k}, \theta_{\text{D},k})$. Therefore, we can summarize the overall mapping of the dequantizer DNN $\mathcal{G}_{\text{D},k}(\cdot)$ with parameter $\mathbf{\Theta}_{\text{D},k}\triangleq\{\mathbf{\theta}_{\text{D},k},\mathbf{C}_{k}\}$ as
\begin{align}
\bar{\mathbf{H}}_{k}=\mathcal{G}_{\text{D},k}(\mathbf{e}_{i_k};\mathbf{\Theta}_{\text{D},k}).\label{eq:deq_NN}
\end{align}

The BS DNN in \eqref{eq:mapping_precoderNN}, developed for processing the continuous-valued actual channel $\mathbf{H}$ handles the dequantized CSI feature $\bar{\mathbf{H}}=[\bar{\mathbf{H}}_1^T,\cdots,\bar{\mathbf{H}}_K^T]^T$. The corresponding output $\mathbf{V}$ is given by
\begin{align}
\mathbf{V} &= \mathcal{G}_{\text{BS}}(\{\bar{\mathbf{H}}_k,\forall k\};\mathbf{\Theta}_{\text{BS}}) \nonumber \\ &=\mathcal{G}_{\text{BS}}(\{\mathcal{G}_{\text{D},k}(\mathbf{e}_{i_k};\mathbf{\Theta}_{\text{D},k}),\forall k\};\mathbf{\Theta}_{\text{BS}}). \label{eq:BS_NN}
\end{align}
Consequently, the end-to-end forwardpass computation $\mathcal{V}:\mathbb{C}^{KN_r \times N_t}\rightarrow\mathbb{C}^{N_t \times KN_r}$ from the CSI $\mathbf{H}$ to the precoding matrix $\mathbf{V}$ is obtained as
\begin{align}
    \mathbf{V} &= \mathcal{V}(\mathbf{H};\Omega) \nonumber \\
    &= \mathcal{G}_{\text{BS}}(\{\mathcal{G}_{\text{D},k}(\mathcal{G}_{\text{G},k}(\mathcal{G}_{\text{P}}(\mathbf{H}_{k};\mathbf{P});\mathbf{\Theta}_{\text{G},k});\mathbf{\Theta}_{\text{D},k}),\forall k\};\mathbf{\Theta}_{\text{BS}}), \label{eq:forward_prop}
\end{align}
where $\Omega$ is defined as
\begin{align}
\Omega \triangleq \big\{\mathbf{P},\{\mathbf{\Theta}_{\text{G},k},\mathbf{\Theta}_{\text{D},k},\forall k\},\mathbf{\Theta}_{\text{BS}}\big\}. \label{eq:Omega}
\end{align}

\subsection{End-to-End Training Policy} \label{sec:training_implementation}
We propose an end-to-end training strategy which integrates the optimization of the pilot NN, the user DNN, and the BS DNN. The multi-stage training policy presented in Sec. \ref{sec:training_PCSI} is adopted to optimize all DNN parameters $\Omega$ in \eqref{eq:Omega} jointly. In addition to the sum-weighted MSE loss \eqref{eq:DNN_sum-weighted MSE}, designing of additional units, e.g., pilot NN and user DNNs, requires a carefully constructed loss function. To this end, we formulate the loss function to consider the operation of the DNN units for CSI acquisition as well as the sum-weighted MSE minimization.

The proposed end-to-end training problem is expressed by
\begin{subequations}
\begin{align}
\min_{\Omega} &\ \ \displaystyle \mathcal{L}(\Omega) + \lambda_{1}\mathcal{R}_{1}(\Omega) +\lambda_{2}\mathcal{R}_{2}(\{\mathbf{C}_{k},\forall k\})\label{eq:DNN_sum_rate_ICSI}\\
\text{subject to }&\ \ \eqref{eq:Ep_constraint},\ \|\mathbf{c}_{k,j}\|^{2}=1, \forall k, j, \label{eq:const}
\end{align}\label{eq:ICSI_task}
\end{subequations}
where the sum-weighted MSE loss function $\mathcal{L}(\Omega)$ is defined as
\begin{align}
\mathcal{L}(\Omega)&\triangleq\mathbb{E}_{\mathbf{H}}\bigg[\sum_{k=1}^{K} \text{Tr}\Big( \mathbf{W}_{\mathbf{\Theta},k}\mathbf{E}_{k}\big(\mathbf{H}_k, \mathcal{V}(\mathbf{H};\Omega)\big)\Big)\bigg],
\end{align}
and the regularization terms $\mathcal{R}_{1}(\Omega)$ and $\mathcal{R}_{2}(\{\mathbf{C}_{k},\forall k\})$ are respectively written by
\begin{align}
\mathcal{R}_{1}(\Omega)&\triangleq\mathbb{E}_{\mathbf{H}}\bigg[\sum_{k=1}^{K} \| \mathbf{H}_k - \mathbf{G}_{k}\|^2_F\bigg]\\
&=\mathbb{E}_{\mathbf{H}}\bigg[\sum_{k=1}^{K} \| \mathbf{H}_k - \mathcal{F}_{\text{G},k}(\mathcal{G}_{\text{P}}(\mathbf{H}_{k};\mathbf{P});\mathbf{\theta}_{\text{G},k})
\|^2_F\bigg], \label{eq:R1}\\
\mathcal{R}_{2}(\{\mathbf{C}_{k},\forall k\})&\triangleq-\mathbb{E}_{\mathbf{H}}\bigg[\sum_{k=1}^{K}\| \bar{\mathbf{G}}_k\mathbf{c}_{k,i_k}\|^2\bigg]. \label{eq:R2}
\end{align}

The contribution of each regularizer in the training loss function \eqref{eq:DNN_sum_rate_ICSI} is controlled by the corresponding nonnegative hyperparameters $\lambda_{1}$ and $\lambda_{2}$. $\mathcal{R}_{1}(\Omega)$ in \eqref{eq:R1} measures the MSE between the actual CSI $\mathbf{H}_{k}$ and the CSI feature $\mathbf{G}_{k}$ in \eqref{eq:Fenc}. It regulates the pilot matrix $\mathbf{P}$ and the user DNN parameters $\mathbf{\Theta}_{\text{G},k}$ such that each user can reconstruct the actual CSI accurately based on the pilot signal $\mathbf{Y}_{k}^{p}$, thereby improving the quality of subsequent processes. Also, $\mathcal{R}_{2}(\{\mathbf{C}_{k},\forall k\})$ in \eqref{eq:R2} focuses only on the codebook regularization by fixing all other trainable parameters. It quantifies the negative affinity between the estimated CSI $\bar{\mathbf{G}}_{k}$ and the associated codeword $\mathbf{c}_{i_{k}}$. Minimizing $\mathcal{R}_{2}(\{\mathbf{C}_{k},\forall k\})$ makes the selected codewords match with the channel matrix and enhances the quantization-dequantization performance by adjusting the quantization points.

Next, we discuss the SGD update rules for solving the constrained training task \eqref{eq:ICSI_task}. At each training epoch, all training parameters in $\Omega$ are first optimized as
\begin{align}
\Omega \leftarrow \Omega - \eta \nabla_{\Omega} \Bigg(\mathbb{E}_{\mathbf{H}}&\left.\bigg[\sum_{k=1}^{K} \text{Tr}\Big( \mathbf{W}_{\mathbf{\Theta},k}\mathbf{E}_{k}\big(\mathbf{H}_k, \mathcal{V}(\mathbf{H};\Omega)\big)\Big)\bigg]\right. \nonumber \\
+\lambda_{1}\mathbb{E}_{\mathbf{H}}\bigg[\sum_{k=1}^{K} &\| \mathbf{H}_k - \mathcal{F}_{\text{G},k}(\mathcal{G}_{\text{P}}(\mathbf{H}_{k};\mathbf{P});\mathbf{\theta}_{\text{G},k}) \|^2_F\bigg]\Bigg). \label{eq:SGD_1st_Tr_ICSI}
\end{align}
Here, $\mathcal{R}_{2}(\{\mathbf{C}_{k},\forall k\})$ is not included in \eqref{eq:SGD_1st_Tr_ICSI} since it is intended only for tuning the codebook $\mathbf{C}_{k}$. The minimization of $\mathcal{R}_{2}(\{\mathbf{C}_{k},\forall k\})$ is thus carried out separately by fixing other parameters~as
\begin{align}
\mathbf{C}_{k} \leftarrow \mathbf{C}_{k} + \eta  \nabla_{\mathbf{C}_{k}}\mathbb{E}_{\mathbf{H}}\bigg[\sum_{j=1}^{K}\| \bar{\mathbf{G}}_j\mathbf{c}_{j,i_k}\|^2\bigg], \forall k \label{eq:cb_update}
\end{align}
where $\lambda_{2}$ is set to one since it does not affect the calculation.
To satisfy the constraints in \eqref{eq:const}, the projected SGD \cite{Kang:18} is applied to the pilot matrix $\mathbf{P}$ and the codebook $\mathbf{C}_{k}$ as
\begin{align}
\mathbf{P} &\leftarrow \sqrt{\frac{T_pE_p}{\text{Tr}(\mathbf{P}\mathbf{P}^H)}}\mathbf{P}, \label{eq:pilot_normalize} \\
\mathbf{c}_{k,j}&\leftarrow \frac{\mathbf{c}_{k,j}}{\Vert \mathbf{c}_{k,j} \Vert_2}, \forall k,j. \label{eq:cb_normalize}
\end{align}

\begin{algorithm}
\caption{End-to-end training policy} \label{algo:end-to-end}
\begin{algorithmic}[1]
\STATE {} Initialize $\Omega$, $\mathbf{P}$ and $\{\mathbf{C}_k ,\forall k\}$.
\STATE {} \textbf{Repeat}
\STATE {} ~~ Sample the mini-batch set $\mathcal{H}_{t}$ from the training dataset.
\STATE {} ~~ Update $\Omega$ in \eqref{eq:SGD_1st_Tr_ICSI} with projections \eqref{eq:pilot_normalize} and \eqref{eq:cb_normalize}.
\STATE {} ~~ Update $\mathbf{C}_k$ in \eqref{eq:cb_update} with projection \eqref{eq:cb_normalize}.
\STATE {} \textbf{Until} convergence
\STATE {} Update $\mathbf{\Theta}_{\text{BS}}$ by using Algorithm \ref{algo:BS_NN_fine}.
\end{algorithmic}
\end{algorithm}

Algorithm \ref{algo:end-to-end} summaries the end-to-end training process which jointly determines the pilot NN, the user DNN, and the BS DNN. We first conduct a joint optimization process which identifies the end-to-end DNN parameter set $\Omega$. At each iteration, all DNN parameters $\Omega$ are updated together using \eqref{eq:SGD_1st_Tr_ICSI} with $\mathbf{W}_{\mathbf{\Theta},k}=\mathbf{I}_{N_r}$. The projection operations \eqref{eq:pilot_normalize} and \eqref{eq:cb_normalize} are employed along with the update of $\Omega$ to implement the projected SGD algorithm. The codebook $\mathbf{C}_{k}$ can be further optimized based on the affinity regularizer $\mathcal{R}_{2}(\{\mathbf{C}_{k},\forall k\})$ in \eqref{eq:R2}. Thus, we fine-tune $\mathbf{C}_{k}$ from \eqref{eq:cb_update} along with the projection \eqref{eq:cb_normalize}. These processes are repeated until $\Omega$ converges. To further improve the training performance, the BS DNN parameter $\mathbf{\Theta}_{\text{BS}}$ can be additionally optimized based on Algorithm \ref{algo:BS_NN_fine}. The optimized BS DNN in $\Omega$ is utilized as the initialization $\mathbf{\Theta}_{\text{BS}}^{\text{init}}$ of the proposed multi-stage training strategy. Through additional multi-stage training, the BS DNN $\mathbf{\Theta}_{\text{BS}}$ can effectively learn the recursive relationship of the WMMSE algorithm \eqref{eq:eq_weight} in the imperfect CSI case. In this multi-stage training step, all other parameters, i.e., the pilot NN and the user DNN, are fixed, since they are not closely related to the precoder optimization. After the offline training, the BS employs the optimized pilot NN parameter $\mathbf{P}$ for conveying the pilot signals as in \eqref{eq:pilotNN}. Then, the users generate the feedback information based on the trained user DNNs \eqref{eq:user_NN}. Finally, the BS computes the precoding matrix by using $\mathbf{\Theta}_{\text{BS}}$. Note that the online implementation of the proposed DNN does not require the perfect CSI $\mathbf{H}$.

\section{Extended Approaches} \label{sec:extension}
This section presents variations of the proposed DL framework to address several practical issues such as the scalability and the channel quality information (CQI) feedback.

\subsection{Scalable Design for Systems with an Arbitrary $K$} \label{section:generalizability}
So far, we have assumed that $K$ for the BS DNN is fixed, but it is desirable to develop a scalable training strategy that can be applied to an arbitrary $K$. Let $K_{\text{max}}$ be the maximum user populations allowed in the system. We first design a DNN according to the maximum population $K_{\text{max}}$. For a scalable architecture, user-specific DNNs such as $\mathcal{G}_{\text{G},k}(\cdot)$ in \eqref{eq:user_NN} and $\mathcal{G}_{\text{D},k}(\cdot)$ in \eqref{eq:deq_NN} share identical trainable parameters, i.e., $\boldsymbol{\Theta}_{\text{G},k}=\boldsymbol{\Theta}_{\text{G}}$ and $\boldsymbol{\Theta}_{\text{D},k}=\boldsymbol{\Theta}_{\text{D}}$, $\forall k$. Also, the input channel realizations $\mathbf{H}\in\mathbb{C}^{K_{\max}N_r \times N_t}$ are constructed according to $K_{\max}$. Then, we uniformly generate active user populations $K\leq K_{\text{max}}$ for each channel training sample. Among total $K_{\max}$ population, $K$ users are regarded as active users scheduled to the BS. The remaining $K_{\max}-K$ users become inactive and are not included in a certain training sample.

The DNNs are readily trained by using Algorithm \ref{algo:end-to-end} along with random populations $K$ as additional training samples. The channel matrices corresponding to inactive users are simply replaced by all-zero matrices in the input of the BS DNN $\bar{\mathbf{H}}\in\mathbb{C}^{K_{\max}N_{r}\times N_{t}}$. Likewise, the precoding matrices of the active users in the output of the BS DNN $\mathbf{V}\in\mathbb{C}^{N_{t}\times K_{\max}N_{r}}$ are exploited in the loss function calculation. Such a random pruning operation can be viewed as a generalization of the dropout layer, which resolves the overfitting problem through model ensembling. This enables the DNNs to learn the end-to-end MU-MIMO systems for an arbitrary $K$ rather than focusing on a specific user population. In the online implementation, the channel input for a certain $K'\leq K_{\max}$ is constructed with a fixed user configuration where the remaining $K_{\max}-K'$ users are inactive. Accordingly, the last $(K_{\max}-K')N_{r}$ columns of the output precoding matrix $\mathbf{V}\in\mathbb{C}^{N_{t}\times K_{\max}N_{r}}$ are discarded.

\subsection{Extension to CQI Feedback} \label{sec:CQI_fb}
The proposed framework presented in Section \ref{sec:ICSI} does not explicitly consider the CQI, e.g., the power of the CSI feature matrix $\mathbf{G}_{k}$. The CQI feedback procedure can be easily integrated with the proposed DL structure by modifying the dequantizer DNN $\mathcal{G}_{\text{D},k}(\cdot;\boldsymbol{\Theta}_{\text{D},k})$, more precisely, the codeword recovery step \eqref{eq:ckik}. To inform the CQI, users can report a quantized version of the channel norm $||\mathbf{G}_k||_F$, denoted by $\mathcal{Q}(||\mathbf{G}_k||_F)$, to the BS. Here, the quantization scheme $\mathcal{Q}(\cdot)$ can be realized by the Lloyd algorithm \cite{Linde:80}. The quantized CQI is multiplied to the codeword in \eqref{eq:ckik} to create a modified codeword $\tilde{\mathbf{c}}_{k,i_{k}}\triangleq\mathcal{Q}(||\mathbf{G}_k||_F)\mathbf{c}_{k,i_{k}}$. Then, the modified codeword is directly utilized as the input feature of the subsequent DNN module. Unlike the unit-norm vector $\mathbf{c}_{k,i_{k}}$, the modified codeword includes the side information about the channel gain. Thus, it is beneficial to handle users with heterogeneous signal attenuations. The effectiveness of this simple modification will be verified in the next section.

\section{Numerical Results} \label{sec:numerical_results}
\begin{table}[]
\caption{DNN Structure According to the Channel Model}
\centering
\begin{tabular}{c|c|c}
\cline{1-3}\hline
\multirow{2}{*}{FC DNN} & \multicolumn{2}{c}{Hidden Layer Dimension} \\ \cline{2-3}
                        &    Rayleigh       &   mmWave        \\ \hline
$\mathcal{F}_{\text{G},k}(\cdot;\mathbf{\theta}_{\text{G},k})$ $(L_{\text{G}}=1)$                       &  $10N_tN_r$         &    100       \\
$\mathcal{F}_{\text{D},k}(\cdot;\mathbf{\theta}_{\text{D},k})$ $(L_{\text{D}}=1)$                       &  $10N_tN_r$         &    100       \\
$\mathcal{F}_{\text{U}}(\cdot;\mathbf{\theta}_{\text{U}})$ $(L_{\text{U}}=3)$                       &  $20N_tN_r$         &    100       \\
$\mathcal{F}_{\text{W}}(\cdot;\mathbf{\theta}_{\text{W}})$ $(L_{\text{W}}=3)$                       &  $40N_tN_r$         &    200       \\ \hline\cline{1-3}
\end{tabular}
\label{table:DNN_setup}
\end{table}
In this section, we assess the performance of the proposed DL based MU-MIMO systems. We first consider the Rayleigh fading channel, whose elements of the channel matrices follow an independent and identically distributed complex Gaussian distribution with zero mean and unit variance. The noise variance at users are set to $\sigma^{2}_k=\sigma^2,\forall k$. The SNR is then defined as $\text{SNR}\triangleq E_s/\sigma^2$. The number of the transmit antennas is set to $N_{t}=KN_{r}$. The structure of the DNNs is presented in Table \ref{table:DNN_setup}. Each hidden layer is followed by batch normalization \cite{Ioffe:15} and employs the rectified linear unit (ReLU) activation function $\text{ReLU}(z)\triangleq\max \{0,z\}$. At the output layer, the identity activation function is applied. For scalable designs, we employ identical DNNs at all users, i.e., $\mathbf{\Theta}_{\text{G},1}=\cdots=\mathbf{\Theta}_{\text{G},K}$ and $\mathbf{\Theta}_{\text{D},1}=\cdots=\mathbf{\Theta}_{\text{D},K}$. The trainable parameters are initialized according to the Xavier initialization method \cite{Glorot:10}, and the Adam optimizer \cite{Kingma:15} is exploited as the mini-batch SGD algorithm. We randomly generate 1000 mini-batch samples at each training epoch, and the DNNs are examined with independent 10000 validation samples. The initialization process to determine $\mathbf{\Theta}^{\text{init}}_{\text{BS}}$ and the first training stage of $\mathbf{\Theta}_{\text{BS}}$ is conducted for 20000 epochs, whereas the subsequent stages are trained for 5000 epochs. During the training of each stage, the learning rate gradually decreases from $\eta=10^{-3}$ to $10^{-4}$. All simulations are implemented with Tensorflow 2.x. 

\subsection{Perfect CSIT} \label{sec:Sim_PCSI}
\begin{figure}
\begin{center}
\includegraphics[width=3.7in]{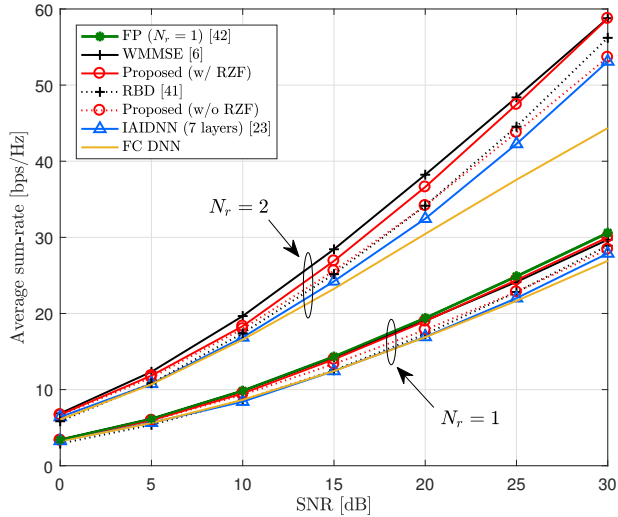}
\end{center}
\caption{Average sum-rate with respect to SNR with $K=4$ and $N_t=KN_r$ in perfect CSIT.}
\label{figure:Sum-rate_PCSI}
\end{figure}
First, we validate the effectiveness of the proposed BS DNN in the perfect CSIT case. Fig. \ref{figure:Sum-rate_PCSI} illustrates the sum-rate performance as a function of the SNR for $N_r=1\text{ \& }2$. As benchmarks, we consider conventional transmitter optimization methods including the WMMSE \cite{Christensen:08} and the regularized block diagonalization (RBD) \cite{Stankovic:08}. The performance of the fractional programming (FP) approach \cite{Shen:18}, which is applicable only for the single-stream case, is included for the single receive antenna case $(N_{r}=1)$. In addition, the following DL-based precoder design schemes are compared to the proposed scheme.
\begin{itemize}
\item \textit{IAIDNN \cite{Hu:20}}: A deep unfolding technique is applied with 7 layers which identifies the MIMO precoding matrix.
\item \textit{FC DNN}: The BS DNN is generated as a naive black-box structure to compute the precoding matrix as an output. The BS adopts a FC DNN consisting of three hidden layers with the dimension $60N_tN_r$, resulting in a similar number of trainable variables as the proposed structure.
\end{itemize}

To investigate the impact of \eqref{eq:RZF}, we present the performance of the proposed method with and without the RZF precoder input. For fair comparison, the DNNs with the same dimension are used for both cases. From the figure, we first see that the proposed DL method achieves the performance quite close to the locally optimum WMMSE algorithm and the FP approach. In particular, it is observed that the performance of our proposed scheme is the same as that of the WMMSE at high SNR. It can be seen that it is beneficial to employ the RZF precoder \eqref{eq:RZF} as the additional input feature. The proposed method is superior to the FC DNN, especially in the MIMO case. This verifies the effectiveness of our model-driven design for the DNN architecture and its training strategy. It is interesting to observe that the proposed method performs better than the IAIDNN which is based on the structure of residual neural network (ResNet) and has shown the best sum-rate performance among existing DL approaches in the literature. These results demonstrate the effectiveness of the proposed learning architecture and its multi-stage training policy.

\begin{figure}
\begin{center}
\includegraphics[width=3in]{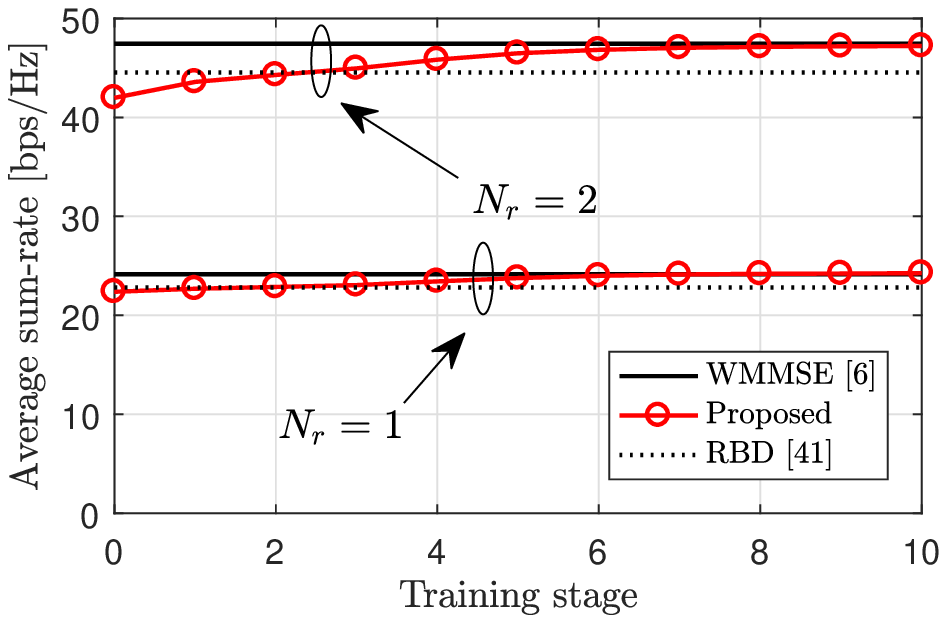}
\end{center}
\caption{Convergence behavior of the proposed DNN approach with $K=4$ and $N_t=KN_r$ at SNR = $25$ dB in perfect CSIT.} \label{figure:convergence_behavior}

\begin{center}
\includegraphics[width=3in]{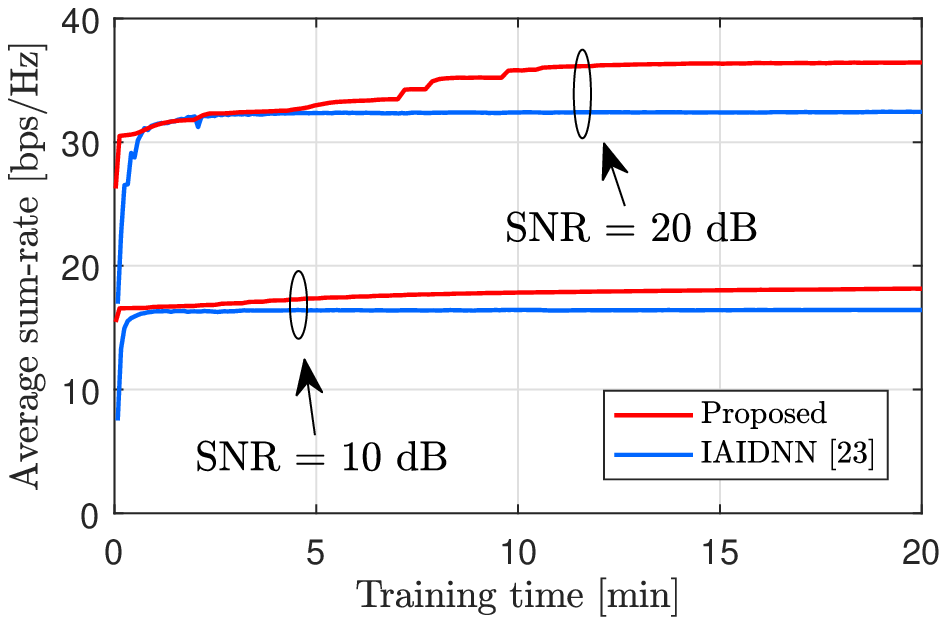}
\end{center}
\caption{Convergence behavior comparison of DNN training with $K=4$, $N_t=8$ and $N_r=2$ in perfect CSIT.} \label{figure:training}
\end{figure}

Fig. \ref{figure:convergence_behavior} evaluates the proposed multi-stage training strategy in terms of the average sum-rate performance at $\text{SNR}=25$ dB with respect to the training stage where the initial training step is denoted as the $0$-th stage. We can observe that our proposed training strategy approaches the conventional WMMSE method \cite{Christensen:08, Shi:11} as the training stage progresses. The plot shows that 6-7 stages are sufficient to achieve the converged performance.

Fig. \ref{figure:training} presents the convergence behavior of the proposed multi-stage training algorithm with respect to the training time. Here, we also examine the performance of the IAIDNN method \cite{Hu:20}. For fair comparison to the method of the IAIDNN, the training complexity is examined with the CPU. The results in Fig. \ref{figure:training} validate the effectiveness of the proposed training policy in Algorithm \ref{algo:BS_NN_fine} for the sum-rate maximization. Moreover, it is observed that the proposed multi-stage training strategy outperforms the IAIDNN method in terms of the final sum-rate performance.

\begin{table}[]
\centering
\caption{Comparison of the Average CPU Running Time in Perfect CSIT with $N_t=KN_r$ and $K=4$ [$\mathrm{msec}$]}\label{table:Time_PCSI}
\subtable[DL approaches]{
\begin{tabular}{|c||c|c|c|}
\hline
                           & FC DNN & Proposed & IAIDNN \\ \hline \hline
\multicolumn{1}{|c||}{$N_r=1$} & 0.048     & 0.279    & 3.734  \\ \hline
\multicolumn{1}{|c||}{$N_r=2$} & 0.634     & 0.589    & 4.549  \\ \hline
\end{tabular} \label{table:Time_PCSI_DL}
}
\subtable[WMMSE]{
\begin{tabular}{|c||c|c|c|}
\hline
                           & SNR = 10 dB & SNR = 20 dB & SNR = 30 dB \\ \hline \hline
\multicolumn{1}{|c||}{$N_r=1$} & 6.917     & 15.501    & 15.676  \\ \hline
\multicolumn{1}{|c||}{$N_r=2$} & 17.179     & 56.112    & 91.436  \\ \hline
\end{tabular} \label{table:Time_PCSI_WMMSE}

}

\subtable[FP]{
\begin{tabular}{|c||c|c|c|}
\hline
                           & SNR = 10 dB & SNR = 20 dB & SNR = 30 dB \\ \hline \hline
\multicolumn{1}{|c||}{$N_r=1$} & $3.587 \times 10^3$     & $3.219 \times 10^3$    & $3.141 \times 10^3$  \\ \hline
\end{tabular} \label{table:Time_PCSI_FP}
}
\end{table}

Table \ref{table:Time_PCSI} compares the CPU running time of various schemes. The layered unfolding architecture of the IAIDNN results in high computational complexity. On the other hand, the proposed approach only employs a single BS DNN module, thereby achieving a comparable time complexity to the FC DNN baseline and the lowest running time among all schemes especially in $N_r=2$. The FP algorithm, which relies on convex solvers at each iteration, shows the worst execution time. The WMMSE algorithm generally requires more iterations at the high SNR regime for convergence, whereas the computations of the DNNs are fixed by the number of the hidden layers and their dimensions regardless of SNR.

\subsection{Imperfect CSIT}

We now investigate a general imperfect CSIT case with $E_{p}=E_{s}$ and $T_p=N_t$. The hyperparameters are set to $\alpha=1.5B$, $\lambda_1=0.1$, and $\lambda_{2}=1$. As a benchmark, we consider the limited feedback system that consists of the LMMSE channel estimator with orthogonal pilot sequences and the codebooks generated by the Lloyd algorithm \cite{Linde:80}. The robust WMMSE algorithm \cite{Fritzsche:13} is also compared, which determines $\mathbf{V}$ based on the channel error statistics. To this end, we numerically measure the error covariance matrices of the overall limited feedback system.

\begin{figure}
\begin{center}
\includegraphics[width=3.7in]{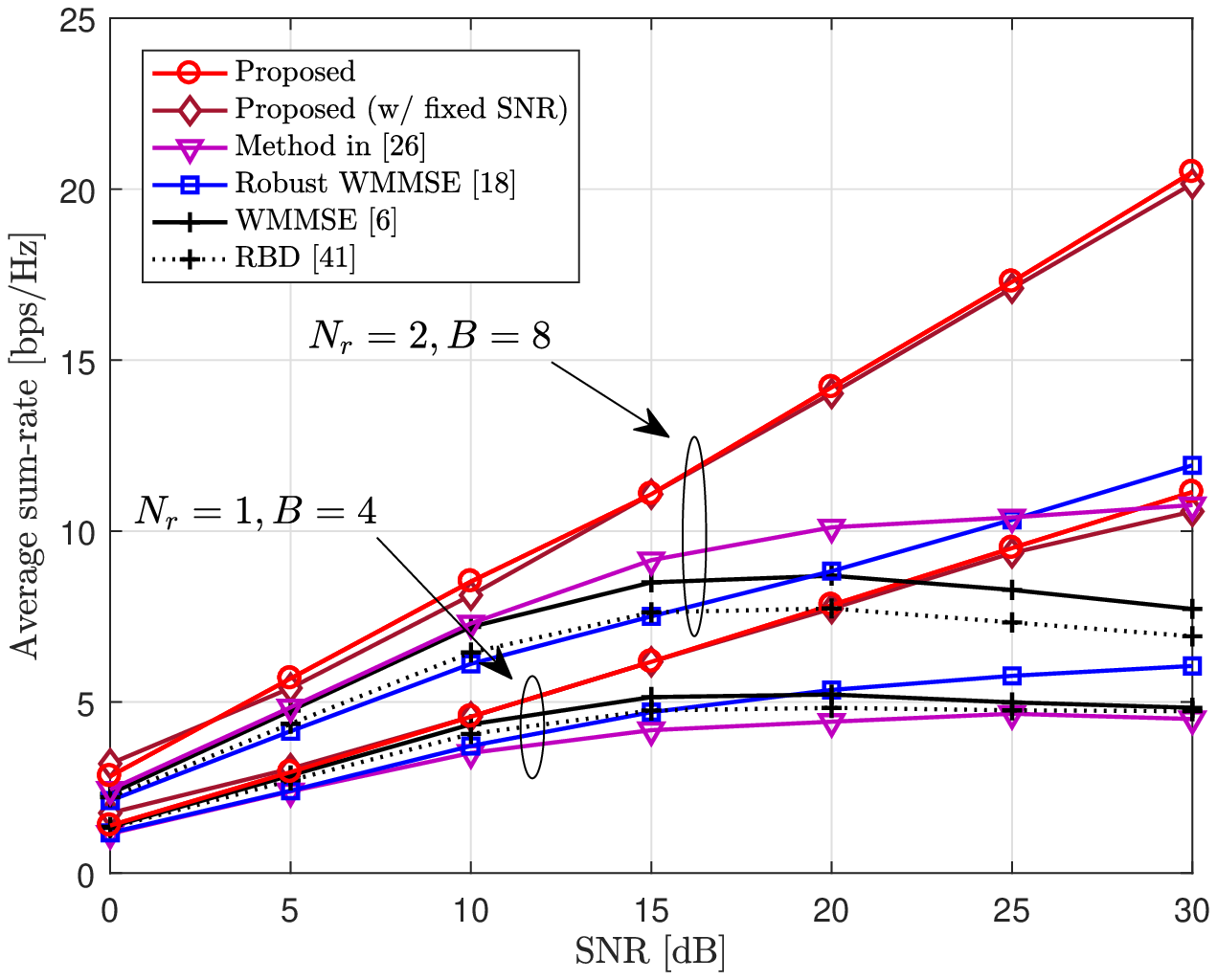}
\end{center}
\caption{Average sum-rate with respect to SNR for $K=4$ and $N_t=T_p=KN_r$.} \label{figure:Sum-rate_FDD}
\begin{center}
\includegraphics[width=3.7in]{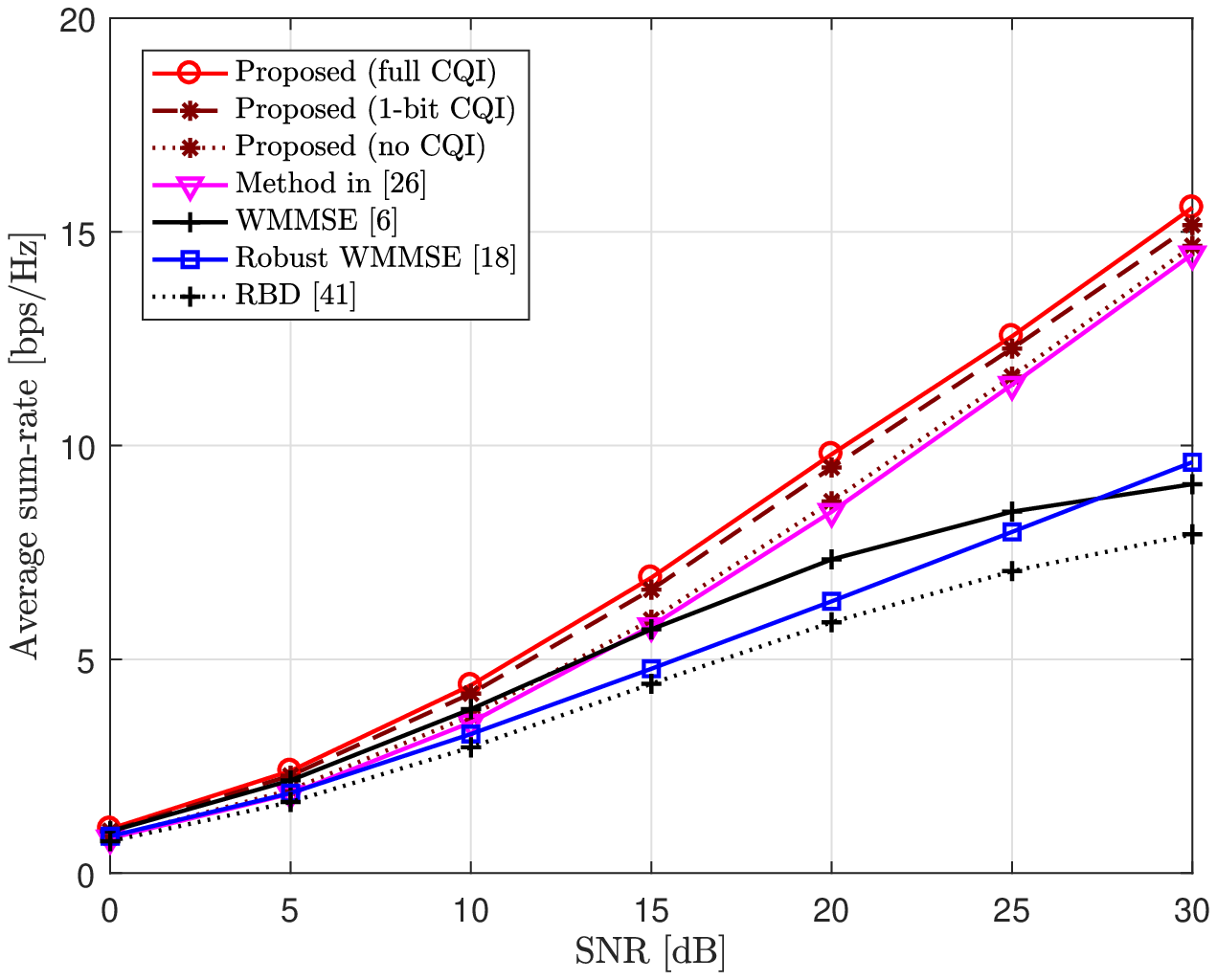}
\end{center}
\caption{Average sum-rate in large scale fading with respect to SNR for $K=2$, $N_t=T_p=KN_r=4$ and $B=5$.} \label{figure:rate_cqi}
\end{figure}

In Fig. \ref{figure:Sum-rate_FDD}, we present the sum-rate performance as a function of the SNR in the imperfect CSIT case. Here, ``Proposed w/ fixed SNR" method indicates the case where the proposed DNN trained at SNR = $15$ dB is directly applied to other SNR values without additional training steps. This scheme shows negligible performance loss at all simulated SNR regime. Therefore, we can conclude that the proposed method trained at particular attenuation environments works well with different SNR. Also, it can be observed that the performance of the RBD and WMMSE is saturated in the high SNR regime. This has been reported in the literature \cite{Jindal:06, Ravindran:08} and such a saturation problem is due to increased multi-user interference incurred by inaccurate channel quantization. In sharp contrast to this, the proposed DL approach shows that the performance monotonically improves. Since the inter-user interference cannot be successfully mitigated due to CSIT impairment, the proposed approach adjusts the transmission strategy by controlling the number of active users to decrease the interference. This advantage of the proposed method allows more flexible power allocation to users, which results in better performance than the existing DL-based method in \cite{Kong:21}.

Fig. \ref{figure:rate_cqi} validates the effectiveness of our proposed scheme which accommodates the CQI feedback scenario, addressed in Section \ref{sec:CQI_fb}. To capture the near-far effect, we take the large-scale fading into account. Users are randomly deployed in a circle cell of radius 100 m, whereas the BS is fixed at the center of the cell. The channel matrix between the BS and the $k$-th user is expressed as $\mathbf{H}_k=\sqrt{\rho_k}\tilde{\mathbf{H}}_k$, where $\rho_k$ and $\tilde{\mathbf{H}}_{k}$ denote the path-loss and the Rayleigh fading matrix, respectively. The path-loss is expressed as $\rho_k=\frac{1}{1+(d_k/d_0)^{\delta}}$ where $d_k$ is the distance between the BS and user $k$, $d_0=30$ m stands for the reference distance, and $\delta=3$ denotes the path-loss exponent. We first train the proposed DNN with the continuous CQI $||\mathbf{G}_k||_F$, and the CQI codebook is then obtained with Lloyd algorithm. We consider different CQI feedback methods. ``Full CQI'' utilizes the exact channel norm $||\mathbf{G}_k||_F$ directly, while in the ``1-bit CQI'' method, users send the 1-bit quantized value of $||\mathbf{G}_k||_F$ in the CQI codebook. We consider total $B=5$ feedback bits both for ``Proposed (no CQI)'' and the method in \cite{Kong:21}, which do not have any explicit CQI feedback procedure. For fair comparison, ``Proposed (1-bit CQI)'' respectively assigns four bits for the direction information and one bit for the CQI. It is observed that ``Proposed (full CQI)" exhibits the best sum-rate performance. The 1-bit quantized CQI provides performance close to the full CQI, albeit at much reduced feedback overhead. On the contrary, the performance of ``Proposed (no CQI)'' and ``Method in \cite{Kong:21}'' is degraded due to the absence of the channel power knowledge at the BS. Therefore, we can conclude that the CQI feedback plays a crucial role in practical MU-MIMO systems.

\begin{figure}
\begin{center}
\includegraphics[width=3.7in]{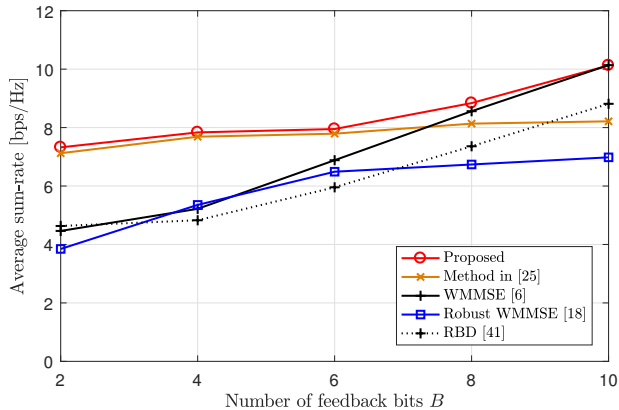}
\end{center}
\caption{Performance comparison with respect to $B$ at SNR $=20$ dB for $K=4$, $N_r=1$ and $N_t=T_p=KN_r$.} \label{figure:rate_B}
\end{figure}

\begin{table}[]
\centering
\caption{Average Sum-rate [$\mathrm{bps/Hz}$] at SNR $=20$ $\mathrm{dB}$ for $K=4$, $N_r=1$ and $N_t=T_p=KN_r$}\label{table:rate_q}
\begin{tabular}{|c||c|c|c|}
\hline
                                                                               & $B=2$   & $B=6$   & $B=10$  \\ \hline \hline
\begin{tabular}[c]{@{}c@{}}Proposed DNN +\\ Proposed quantization\end{tabular} & 7.32 & 7.95 & 10.13 \\ \hline
\begin{tabular}[c]{@{}c@{}}Proposed DNN +\\ Binarization \cite{Sohrabi:21,Kong:21} \end{tabular} & 7.05 & 7.61 & 8.07 \\ \hline
\begin{tabular}[c]{@{}c@{}}Proposed DNN +\\ VQ-VAE \cite{Van:17} \end{tabular} & 7.24 & 7.86 & 9.43 \\ \hline
\end{tabular}
\end{table}

The impact of the feedback bits $B$ at $\text{SNR}=20$ dB in the MU-MISO case is examined in Fig. \ref{figure:rate_B}. The method in \cite{Sohrabi:21}, which designs a DL mechanism for MU-MISO systems, is also compared. The BS DNN in \cite{Sohrabi:21} is intended for the MISO case and is constructed with a FC DNN described in Fig. \ref{figure:Sum-rate_PCSI}. The proposed scheme performs better than other baseline methods in all simulated values of $B$. In particular, a larger performance gain compared to traditional schemes is observed at a small $B$, which results from low rank transmission adjusted by the end-to-end optimization for high channel uncertainty scenarios. The robust WMMSE method, which is designed for the Gaussian channel estimation error, exhibits poor performance at a large $B$ since the acquired channel in the limited feedback system is, in general, interrupted by non-Gaussian errors. This demonstrates that the proposed learning architecture is more suitable for limited feedback systems. A performance improvement over the DL method in \cite{Sohrabi:21} gets larger as $B$ grows.

The effectiveness of the proposed design is examined by comparing with conventional binarization techniques \cite{Sohrabi:21,Kong:21} and the vector quantised-variational autoencoder (VQ-VAE) method \cite{Van:17}. The proposed DNN with VQ-VAE constructs the codewords to minimize the average Euclidean distance from the vectorized CSI feature $\text{vec}(\mathbf{G}_{k})\in\mathbb{C}^{N_{t}N_{r}}$, which requires $N_r$ times larger codeword dimension than the proposed method. In Table \ref{table:rate_q}, the performance of all three methods is shown, and we can see that the proposed approach performs better than the binarization and the VQ-VAE. In addition, a performance gain over the VQ-VAE implies that concentrating on the largest eigenvalue of the training data through \eqref{eq:cb_update} is more efficient in terms of both the compression rate and the precoder performance.

Now, we evaluate the viability of the proposed method in practical millimeter wave (mmWave) channel models presented in \cite{El:14} with uniform linear arrays where azimuth angles are uniformly distributed in $[-30^\circ, 30^\circ]$. The antenna spacing is set to the half wavelength. The channel estimation process of the baseline schemes is conducted by the orthogonal matching pursuit (OMP) method \cite{Tropp:07}. For the limited feedback, the Type-II codebook \cite{3GPP:21} generated with two oversampled beams is investigated. The architecture of the proposed DNN for the mmWave channel is described in Table \ref{table:DNN_setup} with the fixed number of hidden neurons independent of $N_{t}$, $N_r$ and $K$.

\begin{figure} [hbt!]
\begin{center}
\subfigure[Effect of the number of users $K$ ($N_t=32, B=15$)]{
\includegraphics[width=3.1in]{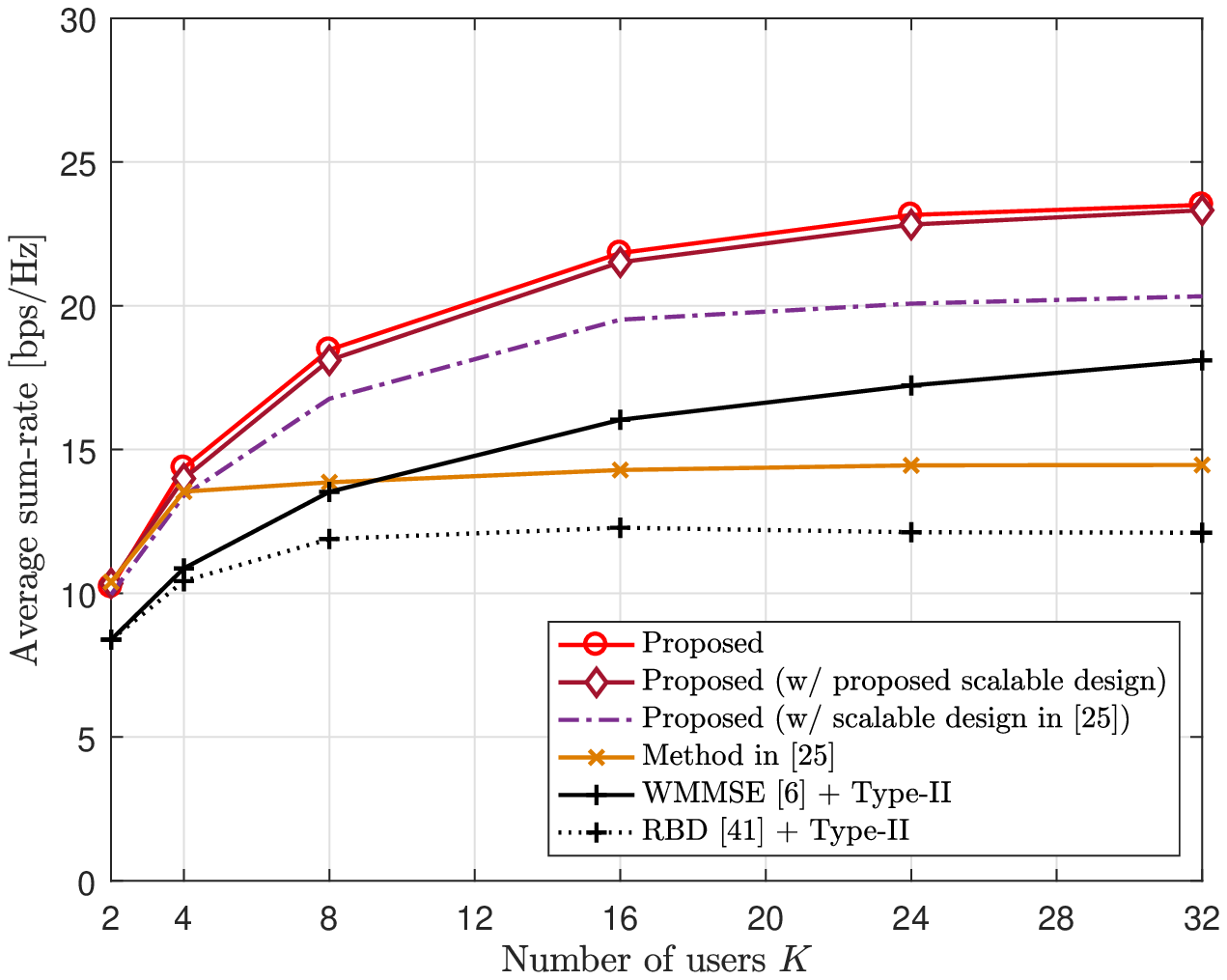}\label{figure:Sum_rate_K}
}
\subfigure[Effect of the number of BS antennas $N_t$ ($B=17, K=8$)]{
\includegraphics[width=3.1in]{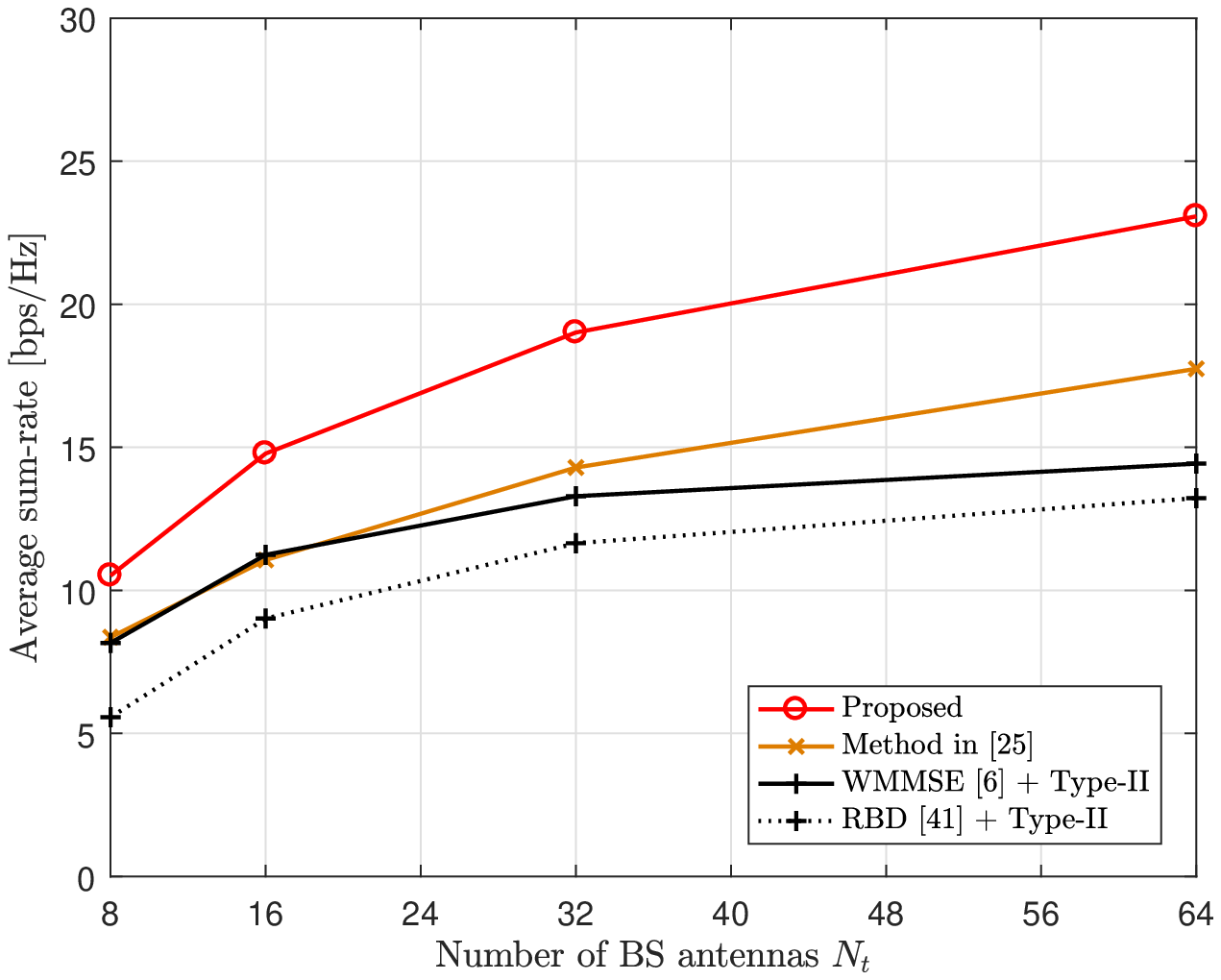}
\label{figure:Sum_rate_Nt}
}
\end{center}
\caption{Average sum-rate in the mmWave channel for $N_r=1$, $E_p=E_s$, $T_p=8$ and $I_p=3$ at $\text{SNR}$ $=10\ \text{dB}$.} \label{figure:Sum_rate_mmWave}
\end{figure}

Denoting $I_p$ as the number of paths in the mmWave channel, Fig. \ref{figure:Sum_rate_mmWave} illustrates the average sum-rate performance at $\text{SNR}=10\ \text{dB}$ for $N_r=1$, $E_p=E_s$, $T_p=8$, and $I_p=3$. We can see that the proposed method clearly outperforms existing methods, which demonstrates that the proposed DNN approach is also powerful in practical mmWave propagation environments. To evaluate a scalable design discussed in Section \ref{section:generalizability}, we investigate the proposed DNN for $K_{\max}=32$ trained with the training data of which each sample is generated with an arbitrary $K\in[24, 32]$, and its performance is illustrated as ``Proposed (w/ scalable design)" in Fig. \ref{figure:Sum_rate_K}. It is interesting to observe that the ``Proposed (w/ scalable design)" scheme shows almost the identical performance to the case individually trained for each $K$, which numerically validates the scalability of the proposed training strategy for an arbitrary $K$. Moreover, we compare the our design to the scalable approach in \cite{Sohrabi:21}, which is denoted as ``Proposed (w/ scalable design in \cite{Sohrabi:21})" scheme. The scalable design in \cite{Sohrabi:21} requires two split training processes. First, the end-to-end DNN is optimized for a virtual single user system, and then the BS DNN is fine-tuned for each given user population $K$. However, such a separate training fails to achieve the end-to-end optimization of all DNN modules. In addition, the method in \cite{Sohrabi:21} needs multiple BS DNNs to calculate the precoding matrices for each given $K$. On the contrary, the proposed scalable design only needs a single DNN for an arbitrary $K$. Furthermore, our method facilitates a joint training of all component DNNs and no additional tuning step is required. In Fig. \ref{figure:Sum_rate_K}, we can see that the proposed scalable design outperforms the method in \cite{Sohrabi:21}, especially in the high $K$ regime. This verifies the superiority of the proposed scalable design which jointly optimizes the user DNN structure for multiple user systems.

Finally, Fig. \ref{figure:Sum_rate_Nt} compares the sum-rate performance of various schemes with respect to $N_t$. It is still observed that our proposed scheme shows superior performance compared to other existing schemes, which demonstrates the suitability of the proposed approach for a large number of transmit antennas.

\section{Concluding Remarks and Future Works} \label{sec:conclusion}
In this paper, we have investigated a novel DL approach for FDD MU-MIMO systems which jointly optimizes the channel acquisition and the transmitter optimization at a BS. To address this challenging problem, multiple NN units have been introduced to determine pilot signals, limited feedback, and precoding. Based on the expert knowledge in the MU-MIMO systems, these NN modules have been carefully designed to maximize the performance via the proposed end-to-end training algorithms. Such an end-to-end training strategy allows the BS and the users to effectively capture any channel impairments incurred in practical MU-MIMO systems. Numerical results have demonstrated that the proposed DL approaches outperform the conventional MU-MIMO techniques with much reduced complexity. As a future work, it is worth pursuing a versatile DL structure that scales up both with antennas and user populations. The proposed scalable design can be extended to train the DNN with the maximum allowable number of antennas. Also, it is important to identify a new DL architecture suitable for the hybrid precoding structure \cite{El:14, Alkhateeb:15}. This invokes a joint optimization of the digital and analog precoding matrices.

\bibliographystyle{ieeetr}
\bibliography{manuscript}

\end{document}